\documentclass[useAMS,usenatbib]{mn2e}
\usepackage{graphicx}
\usepackage{rotating}
\usepackage{lscape}

\title[EFOSC2 Spectroscopy of SWIRE-CDFS galaxies]{EFOSC2 Spectroscopy of SWIRE-CDFS galaxies}
\author[Kalfountzou et al.]
{E. Kalfountzou$^{1,2}$\thanks{Email: e.kalfountzou@herts.ac.uk}, M. Trichas$^{3}$, M. Rowan-Robinson$^{4}$, D. Clements$^{4}$, T. Babbedge$^{4}$ \and J. H. Seiradakis$^{1}$\\
\footnotesize
$^{1}$Aristotle University of Thessaloniki, Dept. of Physics, Section of Astrophysics, Astronomy and Mechanics, GR-541 24\\ Thessaloniki, Greece\\
$^{2}$Centre for Astrophysics, Science \& Technology Research Institute, University of Hertfordshire, Hatfield, Herts, AL10 9AB, UK\\ 
$^{3}$Harvard-Smithsonian Center for Astrophyics, 60 Garden Street, Cambridge MA 02138, USA\\
$^{4}$Astrophysics Group, Imperial College London, Blackett Laboratory, Prince Consort Road, London SW7 2AZ, UK \\}

\begin{document}
\date{Accepted . Received ; in original }

\pagerange{\pageref{firstpage}--\pageref{lastpage}} \pubyear{2010}

\maketitle

\label{firstpage}

\begin{abstract}
We present the optical spectra of a sample of 34 SWIRE-CDFS sources observed with EFOSC2 on the ESO 3.6m Telescope. We have used the spectra and spectroscopic redshifts to validate our photometric redshift codes and SED template fitting methods. 12 of our sources are Infrared Luminous Galaxies. Of these, five belong to the class of ULIRGs and one to the class of HLIRGs with evidence of both an AGN and starburst component contributing to their extreme infrared luminosity for 3, starburst contributing for 1 and AGN contributing for 2 of them.
\end{abstract}


\section{INTRODUCTION}
In recent years, extragalactic surveys have revolutionized our view of the high-redshift Universe, showing that the redshift range $0.5<z<3.0$ witnessed an extraordinary transformation in the demographics of starburst and AGN activity in galaxies. The factors that drove these changes are however unclear. Consequently, the properties of large samples of sources are the current driving force behind modern extragalactic surveys (e.g. Lonsdale et al. 2003; Eales et al. 2010; Oliver et al. 2010). However, either most of the sources in these surveys are too faint for spectroscopic observations or it is not cost and time effective to acquire large spectroscopic redshift samples using the current facilities. Photometric techniques provide an excellent tool for estimating redshifts, essential for statistical studies of the galaxies' properties and evolution with only a fraction of the observing time required by spectroscopic surveys. These techniques are faster and can estimate redshifts beyond the spectroscopic limit with the main handicap of possessing larger errors than spectroscopic techniques. In order to minimize these statistical errors: i) more bands are needed to be included in the photometric computation ii) optical spectroscopic redshifts have to be used to calibrate and validate the photometric redshift codes. \\
In this paper we use the EFOSC2 on the 3.6m ESO telescope to obtain optical spectroscopy of galaxies selected from the SWIRE-CDFS in order to validate our photometric redshift techniques across a wide redshift range, thus allowing more accurate determinations for the whole SWIRE catalogue. In addition, we present the optical spectra and MIR properties of a sample of LIRGs detected as part of our follow-up. Infrared luminous objects at $z>0.5$ are of great importance since they lie at the proposed peak of global starburst activity. They may well be related to the famous sub-mm galaxy population (eg. Chapman et al. 2003) and be connected to the origin of the Cosmic Infrared Background (Puget et al. 1996). Many questions remain about this population, including the role of AGN in their prodigious luminosities, what their descendants are in the local universe, and what their broader role is in galaxy formation and evolution. Some have claimed, for example, that such far-IR luminous objects can only be incorporated into our general understanding of galaxy formation if they have stellar initial mass functions (IMFs) skewed to high masses (Baugh et al. 2005).

\begin{figure*}
\centering
\begin{center}
\includegraphics[scale=0.43]{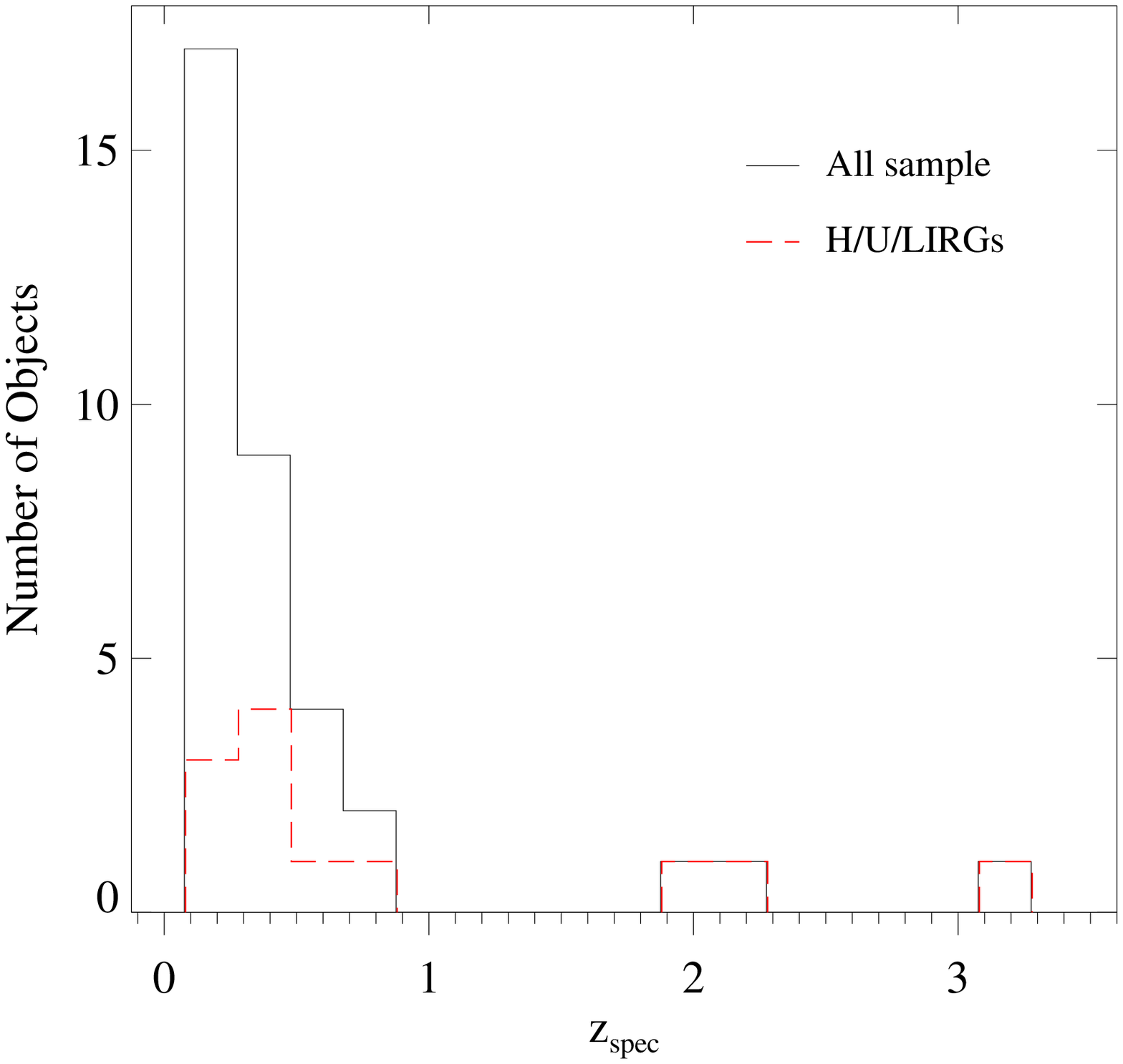}
\includegraphics[scale=0.43]{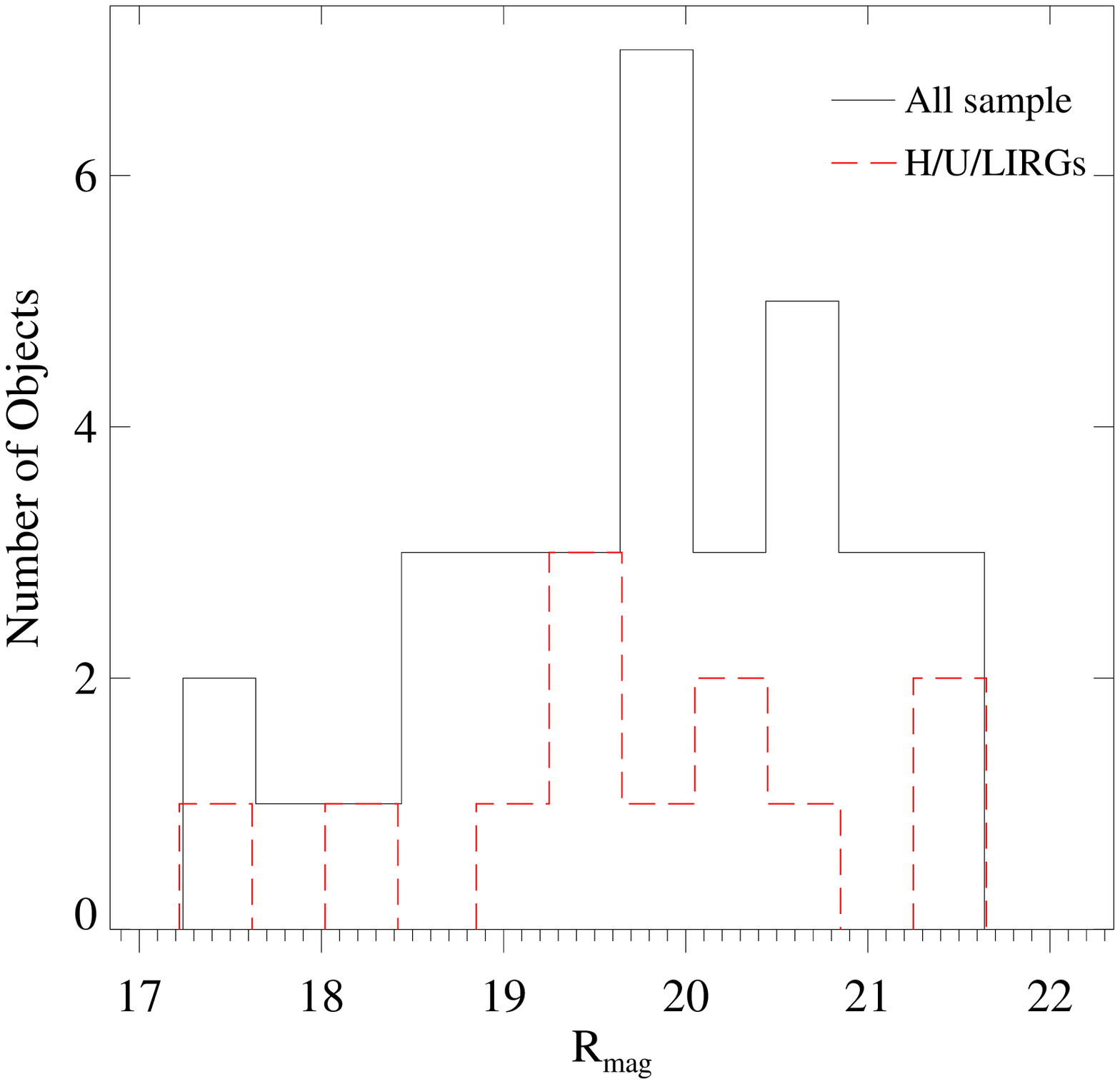}
\end{center}
\caption{\textbf{Left} : The redshift distribution of the 34 extragalactic sources of the spectroscopic sample. \textbf{Rigth} : R-band distribution for the same spectroscopic sample.}
\label{fig:Redshift_distribution}
\end{figure*}

\section{OBSERVATIONS}
The targets for our EFOSC2 run were selected from the available multi-wavelength data in the SWIRE-CDFS field (Lonsdale et al. 2003). Target lists were prepared from the then latest version of the SWIRE Photometric Redshift Catalogue (Rowan-Robinson et al. 2005), selecting 24$\mu$m sources with $r<21.5$. Top priority was given to sources which were of U/LIRG status based on their then estimated photometric redshifts and SED template fits. In total we targeted 12 prime sources in six pointings. The remaining slits were assigned to adjacent serendipitous 24$\mu$m sources. We managed to assign an average of 10 slits per mask, resulting in a total of 62 objects observed. Table 1 gives the list of the six observed pointings and a summary of their properties. \\
The observations were performed in multi-object spectroscopy mode using the ESO Faint Object Spectrograph and Camera (v.2) (EFOSC2) on the ESO 3.6m telescope at La Silla. In order to obtain full coverage over a large wavelength range, both Grism-3 (3050-6100 \AA{}, 1.5 \AA{}/pixel) and Grism-5 (5200-9350\AA{}, 2.06\AA{}/pixel) were used. The width of the slits was 2 arcsec. The observations were carried out during 18-19 December 2006. The exposure time per mask was 2 hours, 3600 seconds per grism. The total time of observations, calibrations, read out time was 15.6h or 2 nights.\\
The EFOSC2 data were reduced and calibrated using standard IRAF routines. The data were bias-subtracted and flat-corrected. The spectra, having been mosaiced, were transformed to a linear wavelength scale. External regions of the 2D spectrum were used to measure the sky background and to subtract it from the final data 1D spectra. Redshifts were determined by visual inspection of the 1D spectra, through identification of emission and absorption features.

\begin{table}
\caption{List of all six EFOSC2 pointing observed. The magnitude refers to the R magnitude of the faintest source in each field.  RA/DEC coordinates refer to the center of each mask.}
\centering
\begin{tabular} [b] {|c| c |c |c|}
\hline
Target-Field 	& $\alpha$(J2000) 	& $\delta$(J2000) 	& Mag. \\
\hline
EA 		& 03h 31m 26s 	& -29d 05m 24s 	& 19.96 \\
EB 		& 03h 35m 29s 	& -28d 45m 00s 	& 21.55 \\
EC 		& 03h 35m 48s 	& -28d 30m 54s 	& 20.60 \\
ED 		& 03h 33m 17s 	& -28d 06m 36s 	& 21.55 \\
EE 		& 03h 28m 55s 	& -28d 46m 12s 	& 18.65 \\
EF 		& 03h 30m 00s 	& -28d 52m 12s 	& 19.11 \\
\hline
\end{tabular}
\end{table}

\begin{table*}
\caption{Properties of the 6 LIRGs, 5 ULIRGs and 1 HLIRG with available spectra from our sample.}
\centering
\begin{minipage}{10in}

\begin{tabular} {| c |c| c |c |c| c | c | c | c | c | c |}
\hline\hline
Object & RA 	& DEC 	& z\footnote{Spectroscopic redshift} & R & $S_\rmn{24}$\footnote{Observed 24$\mu$m fluxes in mJy} &
log($L_\rmn{IR}$)\footnote{Bolometric Infrared Luminosity (1-1000$\mu$m) ($H_\rmn{o}$=72 km $s^{-1}$ $Mpc^{-1}$, $\lambda$ = 0.9)}
& Optical\footnote{Optical SED best fit} & IR\footnote{Infrared SED best fit} & alp2\footnote{Fraction of Starburst contribution at 8$\mu$m}
& alp4\footnote{Fraction of AGN contribution at 8$\mu$m} \\
\hline\hline

SWIRE3-J033523.29-284827.2 & 53.84701 & -28.80756 & 0.195 & 18.61 & 3156.71  & 11.12 & Sbc & M82    & 1    & 0    \\
SWIRE3-J033529.55-284559.2 & 53.87317 & -28.76646 & 0.297 & 17.42 & 0        & 11.20 & Sbc & M82    & 1    & 0    \\
SWIRE3-J033227.95-293122.6 & 53.11647 & -29.52299 & 0.374 & 20.27 & 875.16   & 11.11 & Scd & M82    & 1    & 0    \\
SWIRE3-J033528.00-284500.3 & 53.86671 & -28.75010 & 0.404 & 19.60 & 444.40   & 11.23 & Scd & A220   & 1    & 0    \\
SWIRE3-J033233.17-293004.8 & 53.13818 & -29.50138 & 0.453 & 20.78 & 2920.02  & 11.56 & QSO & A220   & 1    & 0    \\
SWIRE3-J033537.46-284354.0 & 52.90610 & -28.73170 & 0.769 & 21.48 & 700.66   & 11.64 & Scd & M82    & 1    & 0    \\

\hline\hline
SWIRE3-J033234.23-293450.8 & 53.14263 & -29.58084 & 0.529 & 19.80 & 2200.28  & 12.11 & Scd & M82    & 1    & 0    \\
SWIRE3-J033541.23-283414.0 & 53.92179 & -28.57058 & 0.579 & 20.60 & 6248.21  & 12.44 & Scd & M82    & 0.65 & 0.35 \\
SWIRE3-J033532.19-284801.1 & 53.88404 & -28.80030 & 0.807 & 21.55 & 10805.76 & 12.98 & E   & M82    & 0.55 & 0.45 \\
SWIRE3-J033220.88-293140.5 & 53.08703 & -28.52793 & 1.180 & 19.81 & 2486.48  & 12.20 & QSO & Torus  & 0    & 1    \\
SWIRE3-J033528.91-283203.6 & 53.87048 & -28.53434 & 1.962 & 19.32 & 1294.57  & 12.53 & QSO & Torus  & 0    & 1    \\

\hline\hline
SWIRE3-J033144.54-290505.6 & 52.93562 & -29.08486 & 3.200 & 19.96 & 1740.18  & 14.16 & QSO	& M82    & 0.85 & 0.15 \\

\hline\hline
\end{tabular}
\end{minipage}
\end{table*}

\section[]{RESULTS}
Our optical spectroscopic sample contains a total of 62 objects. We have selected to present here the 34 sources of this sample which have a counterpart in the latest version of the SWIRE photoz catalogue (Rowan-Robinson et al. 2008) and their redshift determination is secure. Sources with very noisy or bad quality spectra have been rejected. The spectra of all the 34 sources are shown in Figures 3, 4, 10 and 11 and their properties are shown in Table 4. \\
Figure 1 (left) shows the redshift distribution of the spectroscopic sample. The bulk (85.7\%) of the 34 sources targeted, lie at $z<0.6$. The maximum peak in the redshift distribution  is found at $z<0.2$. The source with the highest redshift found is at z$\sim$3.20. Sources at $z>2$ have prominent Lya lines. In the same figure we show the redshift distribution of the Infrared Luminous Galaxies in our sample. These galaxies correspond to the higher redshift sources of our sample and their main bulk lie at $0.4<z<0.6$. Figure 1 (right) shows the R-band distribution of the 35 extragalactic sources observed. \\
We have used Rowan-Robinson et al. 2008 template fitting method to estimate the infrared luminosities of the 35 observed sources using the observed spectroscopic redshifts. Based on the library of infrared templates used, 11 out of 34 sources ($\sim$32.3\%) are fitted with a cirrus, 13 ($\sim$38.2\%) are fitted with a starburst, 6 are fitted with a dust torus ($\sim$17.6\%) and the rest of the sources have single band infrared excess. According to the optical templates 5 sources are fitted with a QSO template. Of the remaining 29 galaxies, 4 ($\sim$13.3\%) are classified as ellipticals and 25 ($\sim$86.7\%) as spirals. \\
According to the estimated bolometric infrared luminosities, among the 34 sources there are 6 luminous infrared galaxies (LIRGs) ($L_{\rmn{IR}}>10^{11}L_{\rmn{\odot}}$), 5 ultraluminous infrared galaxies (ULIRGs) ($L_{\rmn{IR}}>10^{12}L_{\rmn{\odot}}$) and 1 hyperluminous  infrared galaxy (HLIRGs) ($L_{\rmn{IR}}>10^{13}L_{\rmn{\odot}}$). Table 2 summarizes the main properties of these IR-luminous objects and Figure 3 shows their spectra. Seven of the 12 H/U/LIRGs are fitted with a spiral optical template, 1 is fitted with an elliptical template and 4 are fitted with a QSO optical template (1 LIRG, 2 ULIRGs and 1 HLIRG). According to the infrared templates, 2 ULIRGs are fitted with a dust torus, 8 sources are fitted with an M82 starburst (4 LIRGs, 3 ULIRGs and 1 HLIRG) and 2 LIRGs appear to be A220 starburst fitted. Two of the IR-luminous sources seem to contain a powerful AGN, 7 seem to be powered by a starburst, and the rest 3 seem to be powered both by a starburst and an AGN. Half of the IR-luminous sources with ($L_{\rmn{IR}}>10^{12}L_{\rmn{\odot}}$) are powered both by a starburst and an AGN. These results are consistent with previous studies of 24$\mu$m selected ULIRGs (Trichas et al. 2009, 2010). The only HLIRG of our sample is powered both by a starburst and an AGN with a major contribution of a starburst. Below we summarize the properties of the 5 ULIRGs and 1 HLIRG found. The U/HLIRGs spectra are given in Figure 3.\\
\textbf{SWIRE3-J033234.23-293450.8}: It is one of the strongest 24$\mu$m emitters among all H/ULIRGs of our sample and also the one with the lowest redshift ($z=0.529$). It has been detected in all IRAC and MIPS bands. Balmer lines, [\textit{OII}] and  [\textit{OIII}] doublet are detected at its spectra. Both optical and IR SED fitting imply the presence of a starburst. Starburst's presence is also judged by the emission lines diagnostic diagrams (Figure 5).\\
\textbf{SWIRE3-J033541.23-283414.0}: The R-band magnitude of this source is $R=20.60$. It is a strong 24$\mu$m emitter with detections in all IRAC and MIPS bands. As it can be seen by its spectra there is a very strong [\textit{OII}] line. Very strong [\textit{OIII}] doublet and \textit{H$\beta$} are clearly visible. These narrow lines, based on the results of BPT diagrams, indicate the presence of a starburst. The detection of a quite broad \textit{H$\delta$} with a \textit{MgII} line confirms the presence of an AGN. Both optical and IR SED fitting imply the presence of a starburst but with AGN contribution. The redshift of this sources is $z=0.579$.\\
\textbf{SWIRE3-J033532.19-284801.1}: The R-band magnitude of this source is $R=21.55$. It is the strongest 24$\mu$m emitter of our sample with detections in all IRAC and MIPS bands. As it can be seen by its spectra there is a strong [\textit{OII}] line. \textit{H$\beta$} and [\textit{OIII}] doublet lines are also visible but weaker. A weak and broad \textit{MgII} line is also visible at the blue spectrum. These lines indicate the presence AGN events. The source lies at $z = 0.807$. Optical SED fitting imply the presence of a young elliptical galaxy and the IR SED fitting an AGN dust torus.\\
\textbf{SWIRE3-J033220.88-293140.5}: It is the strongest 24$\mu$m emitter among all 5 ULIRGs, with detections in all IRAC bands and at 24$\mu$m. The R-band magnitude of this source is $R=19.81$. We have detected 2 strong broad lines \textit{CIII} and \textit{MgII}. The presence of these lines makes it almost certain the presence of a very strong AGN which is also identified by optical and IR SED fitting. This source lies at $z = 1.180$.\\
\textbf{SWIRE3-J033528.91-283203.6}: This ULIRG is the brightest one of our sample with R-band magnitude $R=19.32$. It has been detected in all IRAC bands and at 24$\mu$m. As it can be seen by its spectra there are 3 strong and broad detected lines \textit{Ly$\alpha$}, \textit{CIV} and \textit{CIII}. Both optical and IR SED fitting imply the presence of an AGN with significant starburst contribution. The redshift of this sources is $z=1.962$.\\
\textbf{SWIRE3-J033144.54-290505.6}: This is the only HLIRG in our sample with bolometric infrared luminosity $L_{\rmn{IR}}=10^{14.16}L_{\rmn{\odot}}$. This is one of the brightest ($R=19.96$) sources at the highest redshift ($z=3.2$) in this sample. It has been detected in all IRAC and MIPS bands except 160$\mu$m. This source exhibits 4 broad lines \textit{Ly$\alpha$}, \textit{NV}, \textit{SiIV} and \textit{CIV}. The clear detection of these lines indicates the presence of a QSO in agreement with SED fitting. \\

\begin{figure}
\begin{center}
\includegraphics[scale=0.43]{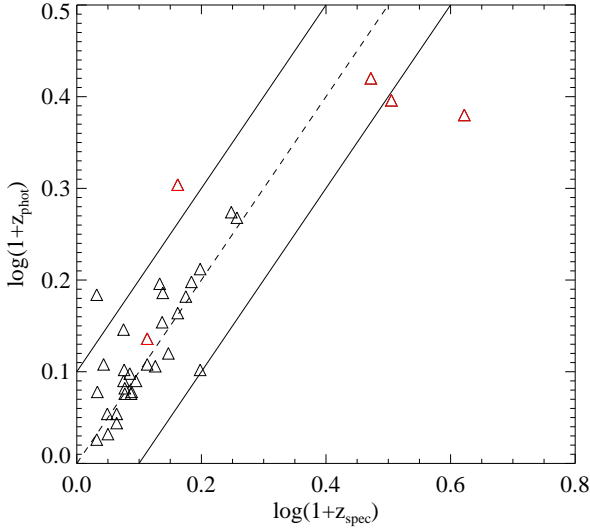}
\end{center}

\caption{Photometric versus spectroscopic redshift for all sources with available spectroscopic redshifts in our sample. The straight lines represent a 10\% accuracy in $log_{\rmn{10}}$(1+z). Red triangles are sources fitted with a QSO template and black triangles are sources fitted with a galaxy template.}
\end{figure}

\begin{figure*}
\begin{center}

\includegraphics[scale=0.55]{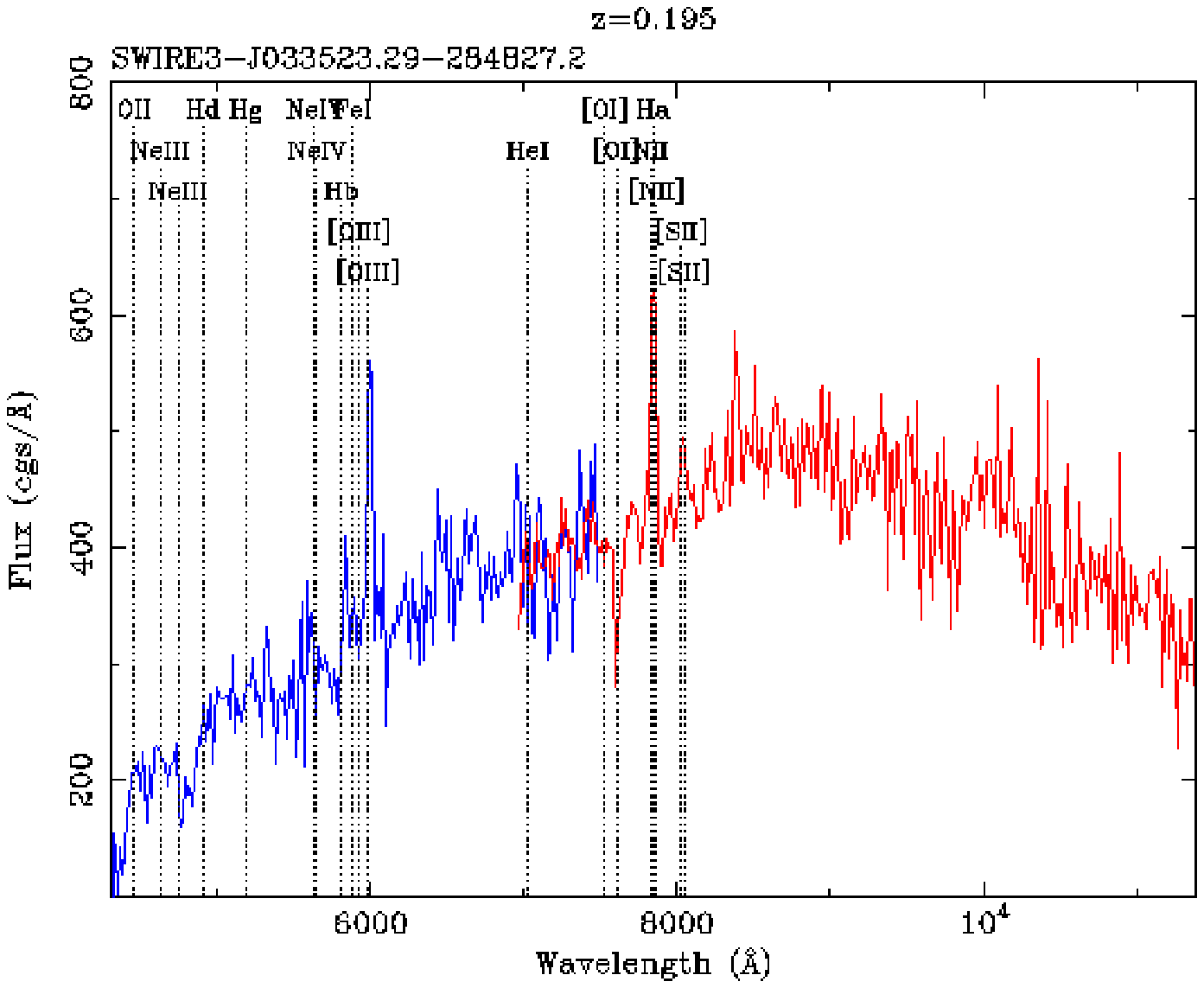}
\includegraphics[scale=0.4]{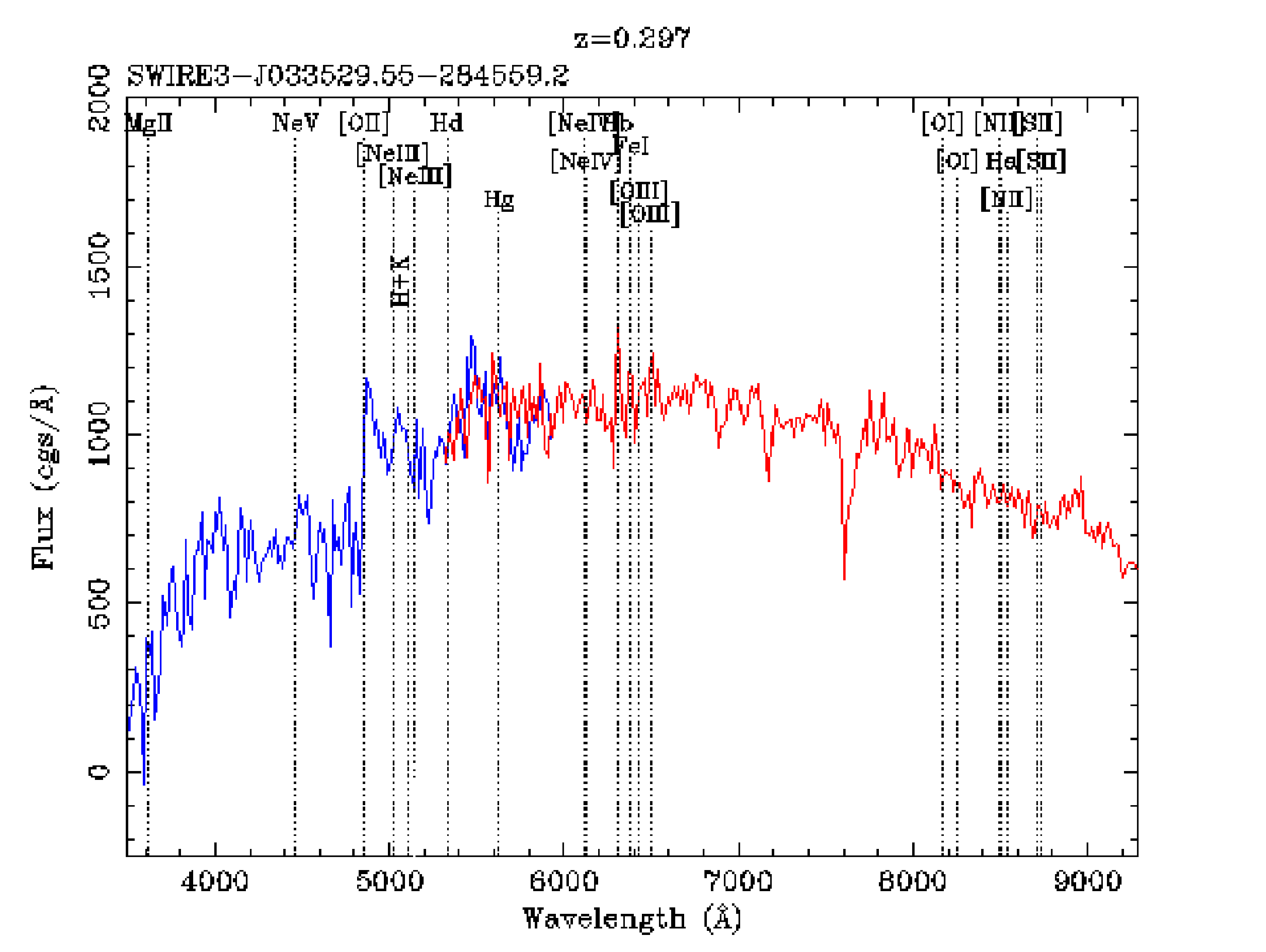}

\vspace{5pt}

\includegraphics[scale=0.55]{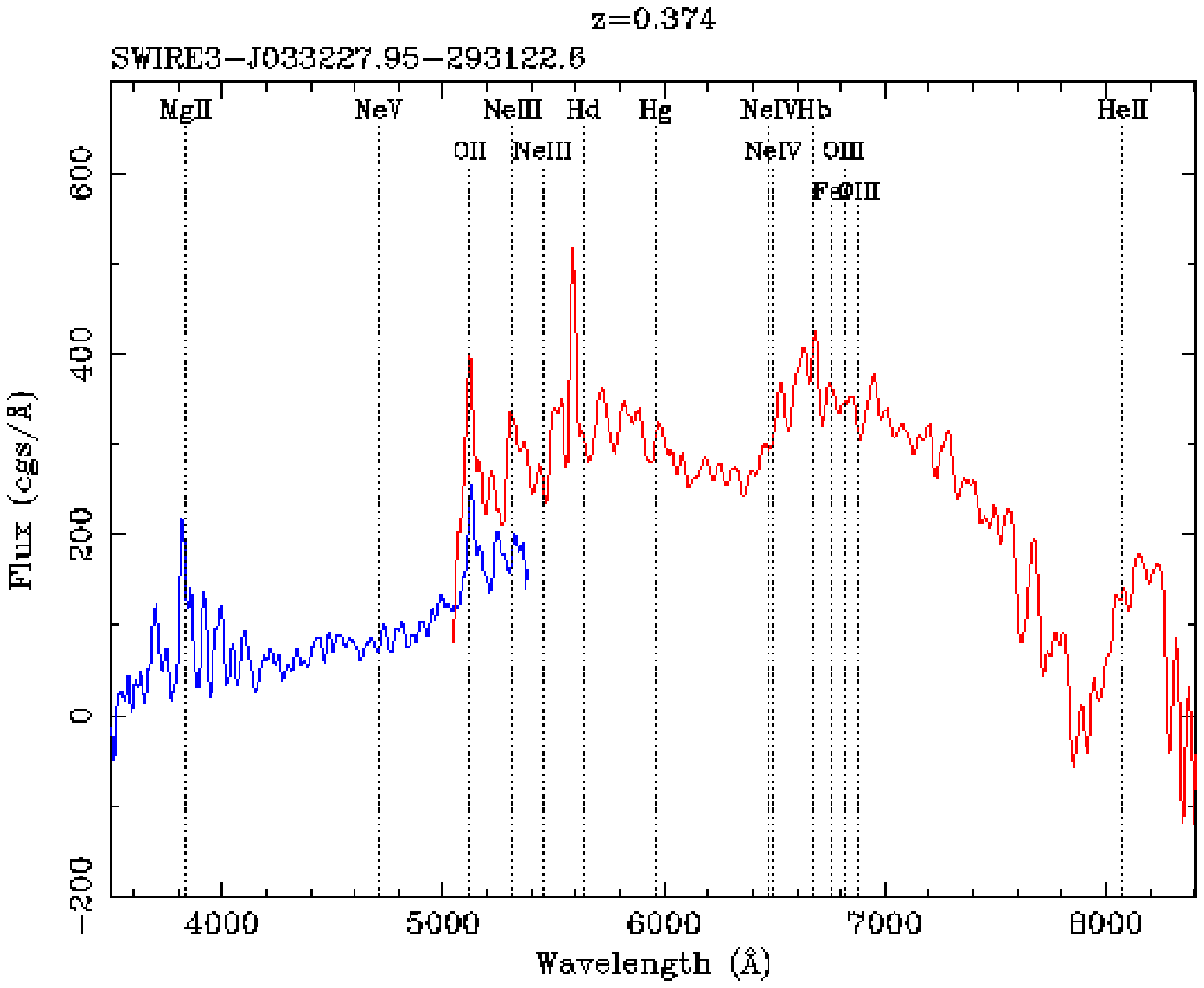}
\includegraphics[scale=0.55]{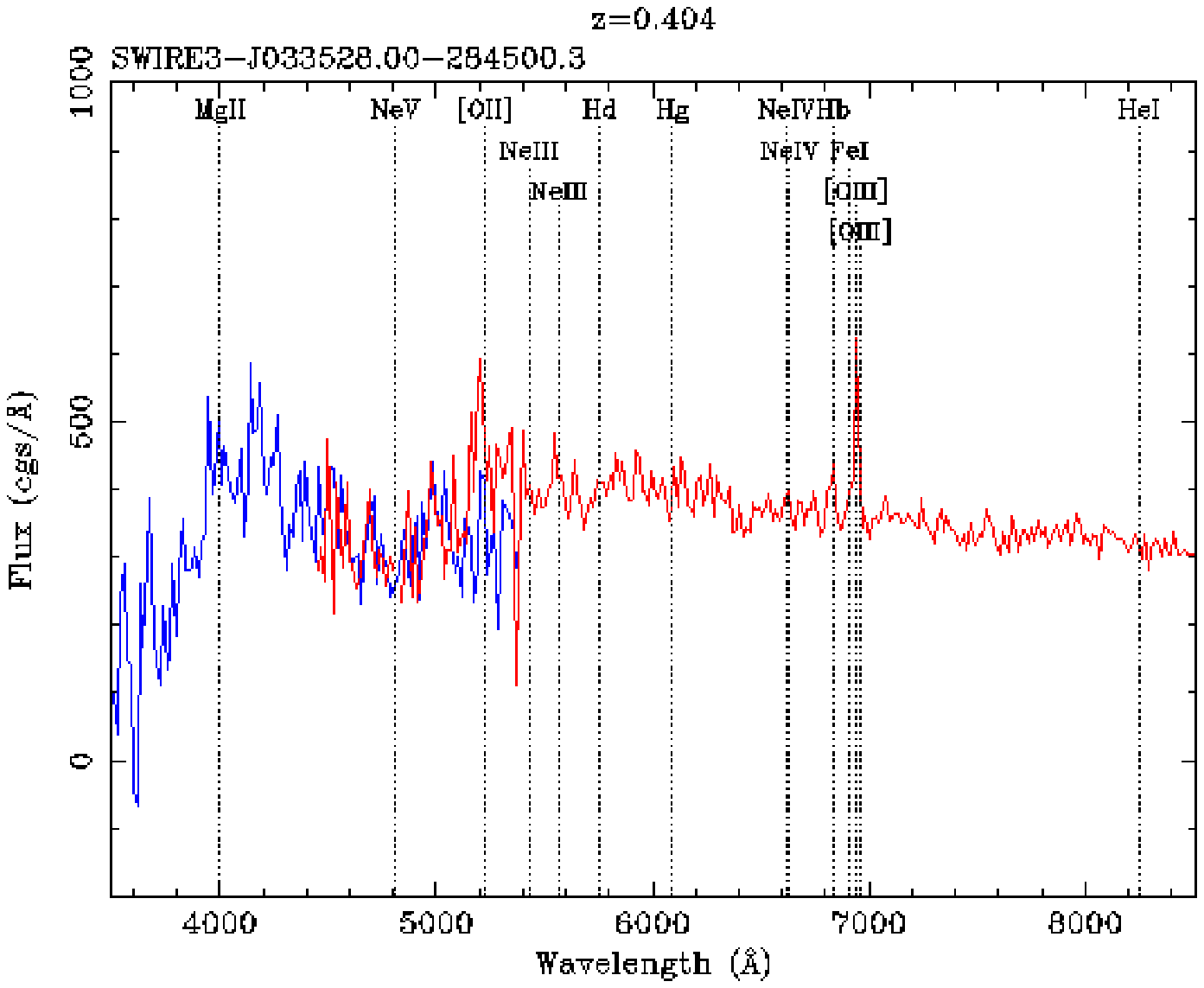}

\vspace{5pt}

\includegraphics[scale=0.55]{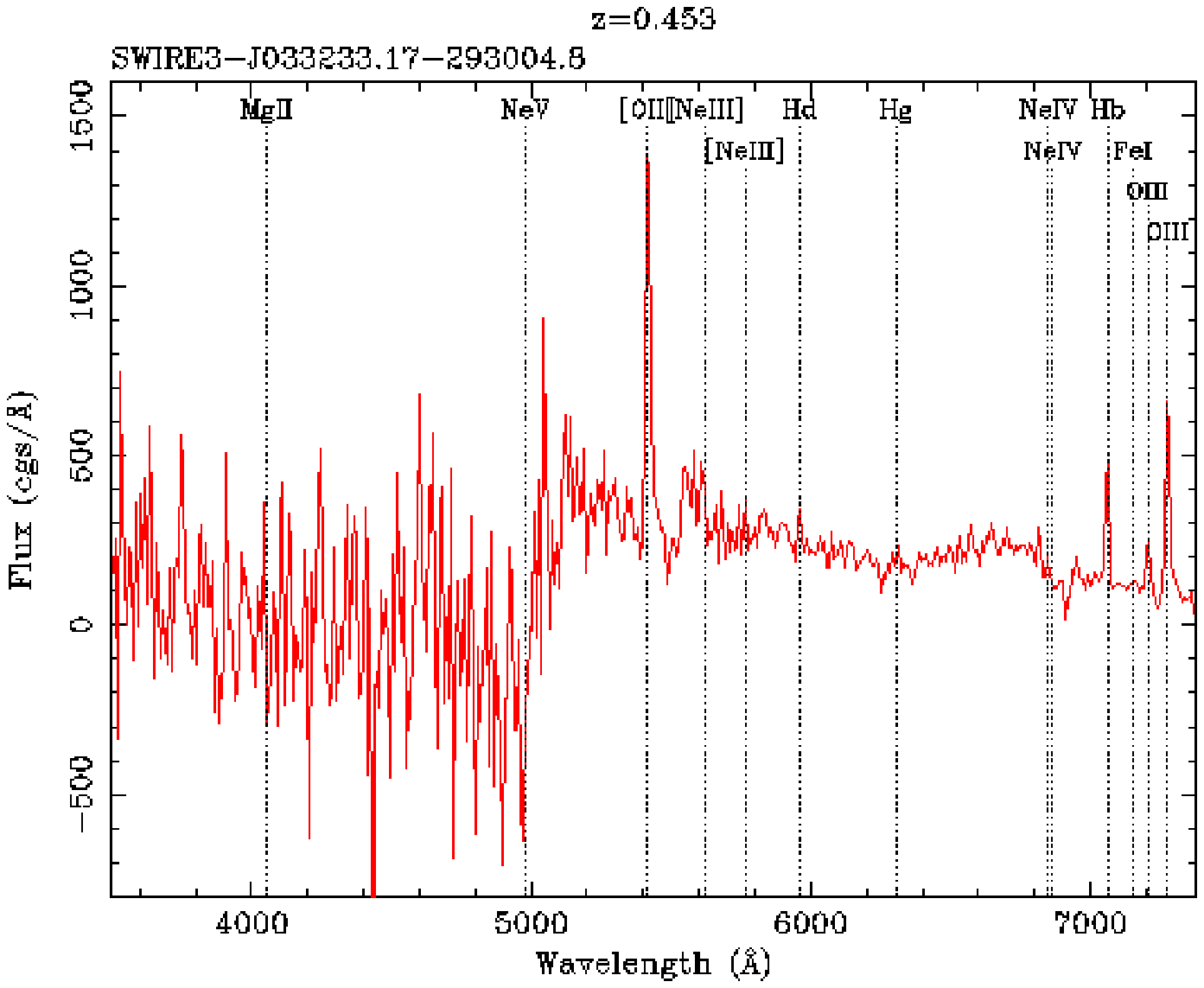}
\includegraphics[scale=0.55]{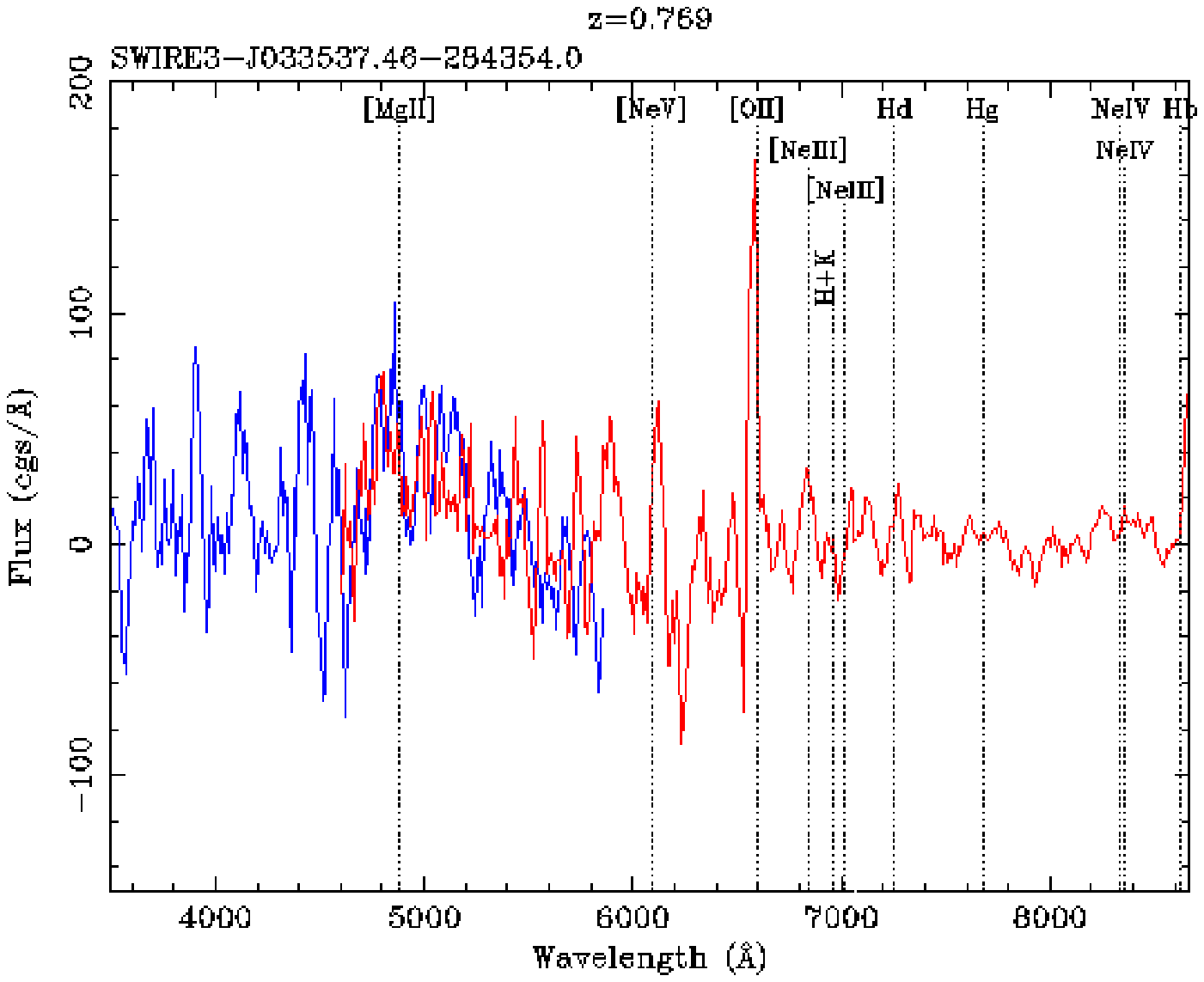}

\end{center}
\caption{Optical spectra of the 6 Luminous Infrared Galaxies found in our sample. Blue and red colors represent spectra taken with Grism-3 and 5 respectively.}
\end{figure*}

\begin{figure*}
\begin{center}

\includegraphics[scale=0.55]{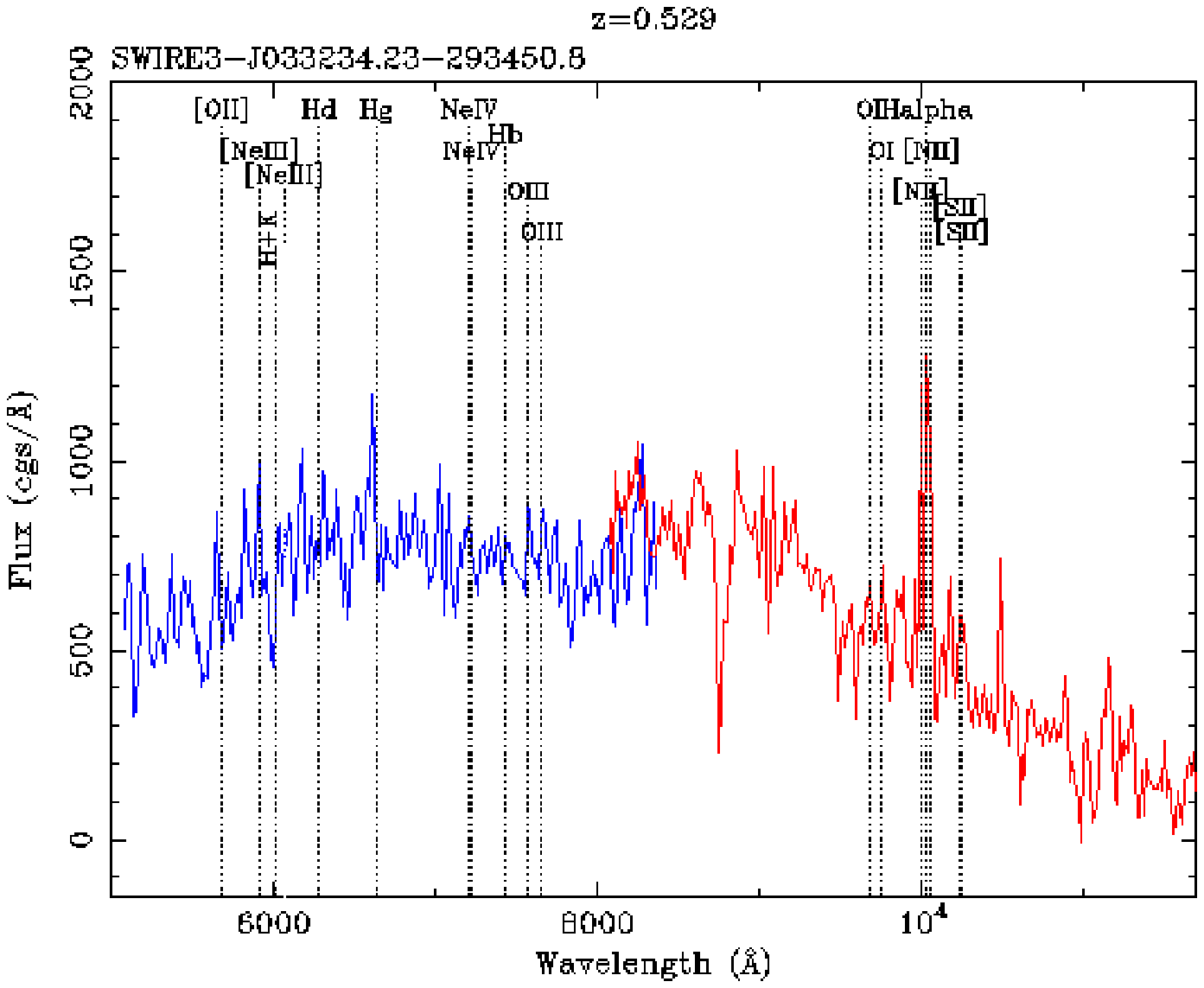}
\includegraphics[scale=0.55]{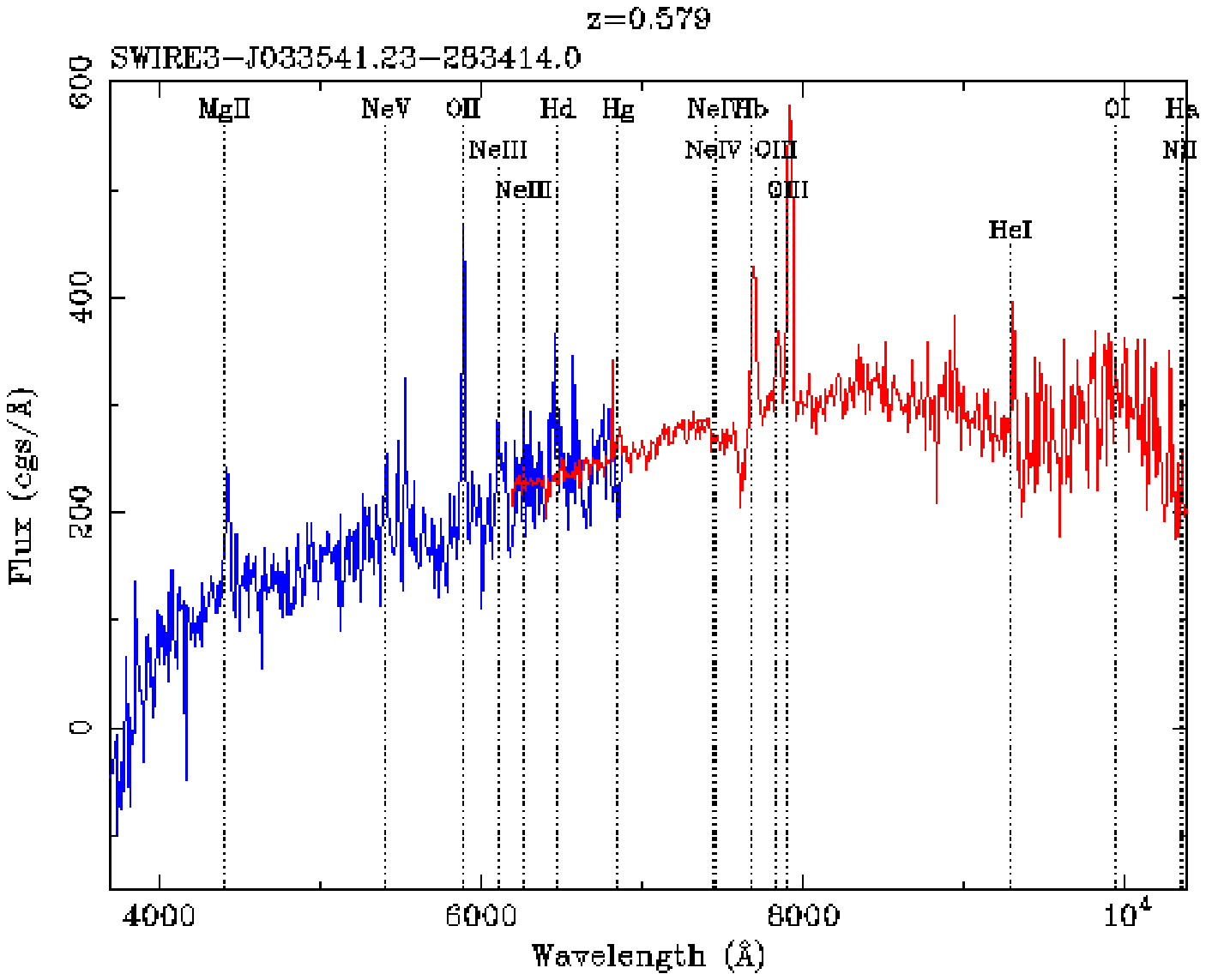}

\vspace{5pt}

\includegraphics[scale=0.55]{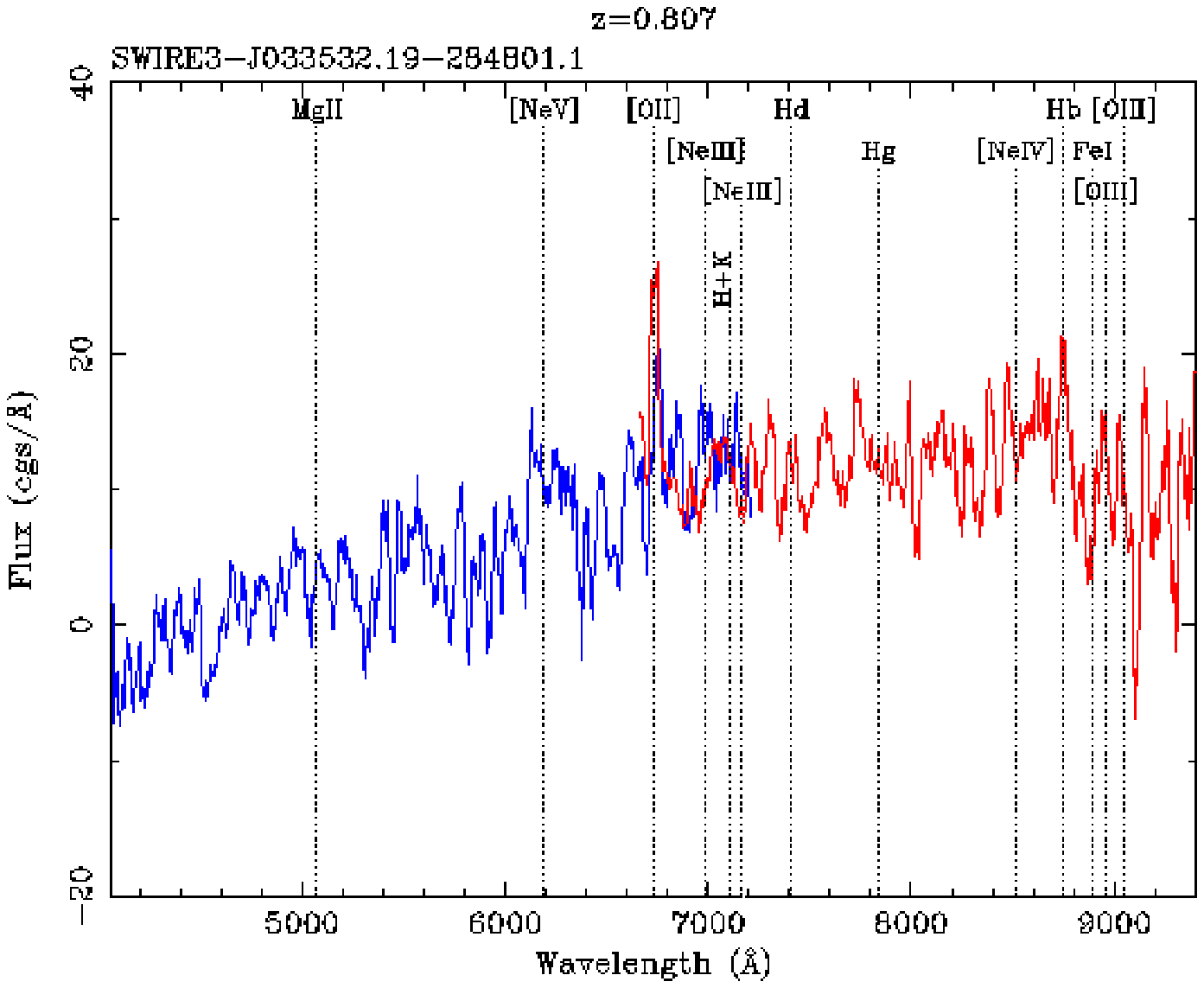}
\includegraphics[scale=0.55]{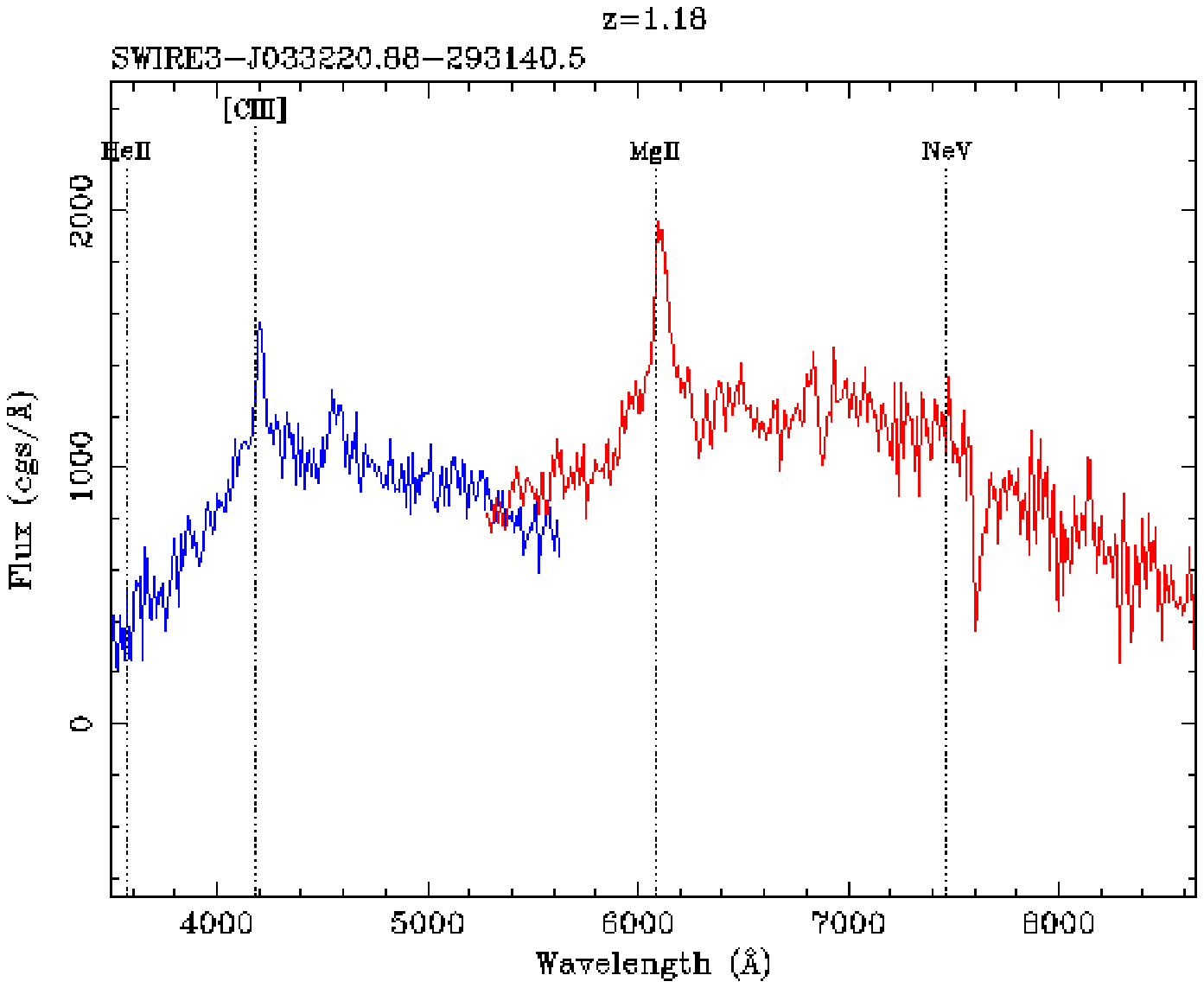}

\vspace{5pt}

\includegraphics[scale=0.55]{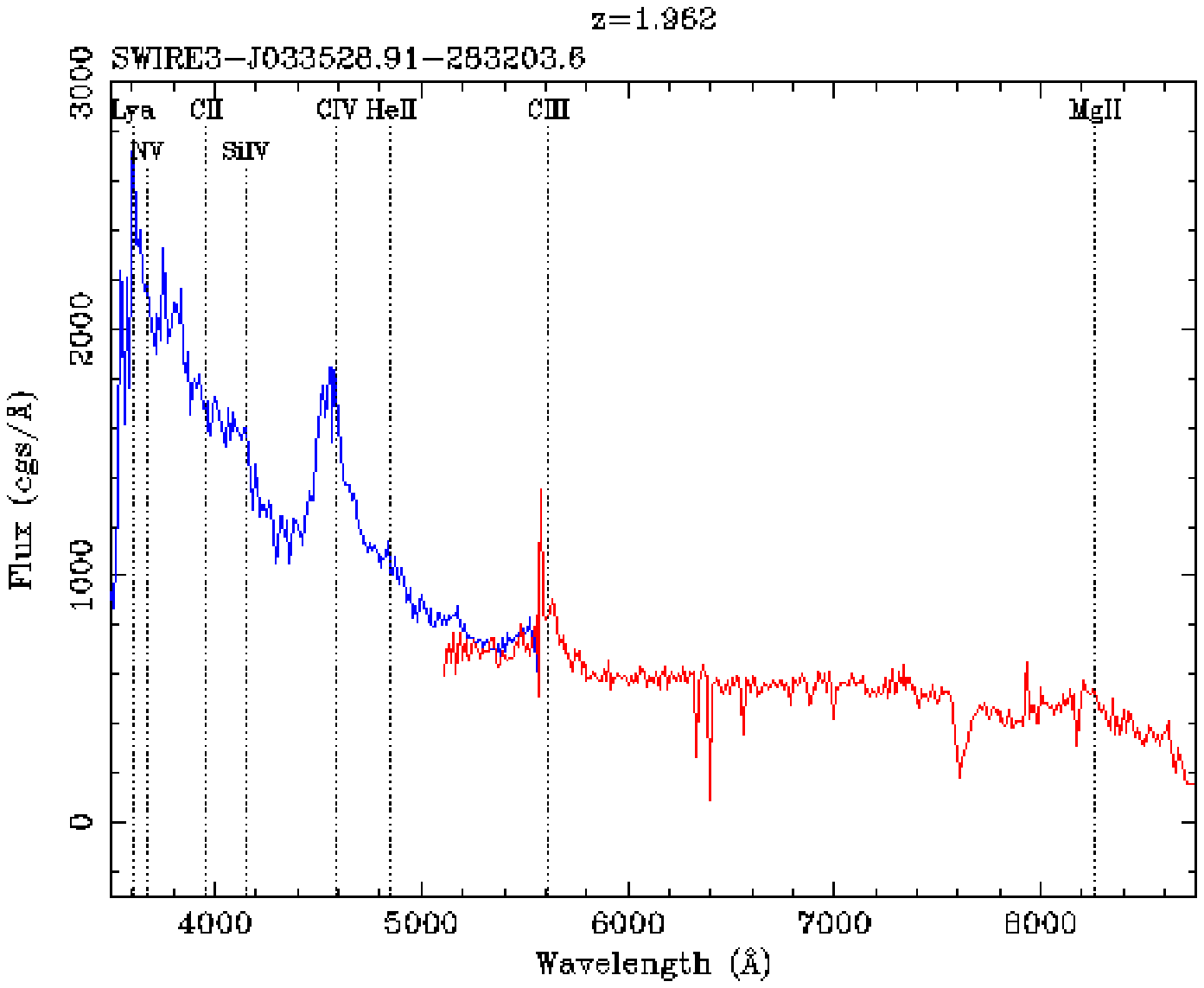}
\includegraphics[scale=0.55]{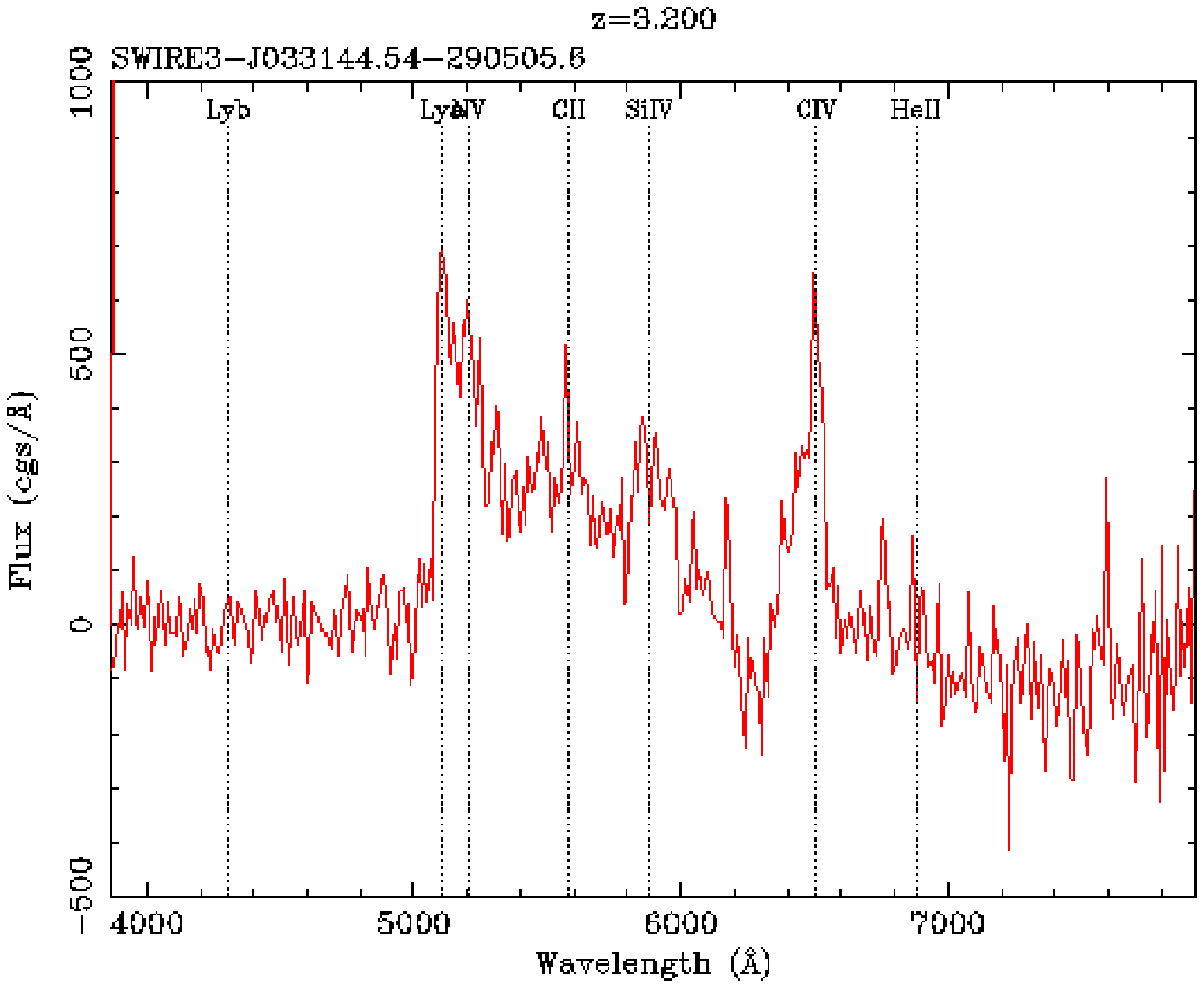}

\end{center}
\caption{Optical spectra of the 5 UltraLuminous and the 1 HyperLuminous Infrared Galaxies found in our sample. Blue and red colors represent spectra taken with Grism-3 and 5 respectively.}
\end{figure*}

\subsection[]{PHOTZ/SPECZ COMPARISON}

We have compared the spectroscopic redshifts obtained with EFOSC2 with the photometric redshifts calculated using the latest version of the ImpZ code (Rowan-Robinson et al. 2008). The results are given in Figure 2. Reliability and accuracy of the photometric redshifts are measured via the fractional error $\Delta$z/(1+z) for each source, examining the mean error $\overline{\Delta}$z/(1+z), the \textit{rms} scatter z and the rate of 'catastrophic' outliers $\eta$, defined as the fraction of the full sample that has \textbar$\Delta$z/(1+z)\textbar$>$0.15. Figure 2 shows a comparison of $log_{\rmn{10}}(1+z_{\rmn{phot}})$ versus $log_{\rmn{10}}(1+z_{\rmn{spec}})$ for the sample of 34 sources with available spectroscopy.  For the whole sample, the total \textit{rms} scatter, $\sigma_{\rmn{z}}$, is 0.121, with $\overline{\Delta}$z/(1+z) = 0.013 and the number of outliers sources is 4. 3/4 of the outliers are fitted with a QSO template and one with a galaxy template. Redshift estimation for all the outliers is based on the detection of at least 3 lines making their specz highly reliable.  From the 4 sources which were fitted with a QSO optical template the number of outliers is 3. From the 3 sources with $z_{\rmn{spec}}>1.5$, all of them fitted with a QSO optical template, there are two "catastrophic" outliers.

\begin{table*}
\caption{The [\textit{NII}]/\textit{H$\alpha$} vs [\textit{OIII}]/\textit{H$\beta$} and [\textit{SII}]/\textit{H$\alpha$} vs [\textit{OIII}]/\textit{H$\beta$} diagrams for all the 15 sources with available lines.}
\begin{minipage}{5.5in}
\begin{tabular} {|c|c| ccc| c | c |}

\hline\hline

Object & z\footnote{Spectroscopic redshift}  & \multicolumn{3}{c}{Measured line ratios}	    &	SC1\footnote{Spectroscopic Classification according to the [\textit{NII}]/\textit{H$\alpha$} vs [\textit{OIII}]/\textit{H$\beta$} diagnostic} &  SC2\footnote{Spectroscopic Classification according to the [\textit{SII}]/\textit{H$\alpha$} vs [\textit{OIII}]/\textit{H$\beta$} diagnostic} 	\\
& & S1\footnote{$log_{\rmn{10}}$([\textit{NII}]/\textit{H$\alpha$})} & S2\footnote{$log_{\rmn{10}}$([\textit{SII}]/\textit{H$\alpha$})} & S3\footnote{$log_{\rmn{10}}$([\textit{OIII}]/\textit{H$\beta$})} &	 & \\

\hline\hline

SWIRE3-J033152.82-290647.1  & 0.079	& -0.8240	&	-0.1677	&	-0.0962     &	Star-forming	&	LINER		\\
SWIRE3-J033148.59-290643.1  & 0.119	& -0.2180	&	-0.5677	&	 0.7244	    &	AGN      	&	AGN	 	\\
SWIRE3-J033145.66-290728.0  & 0.245	& -0.0971	&	-0.1584	&	-0.1121     &	Composite	&	LINER	 	\\
SWIRE3-J033142.11-290551.9  & 0.158	& -0.2765	&	-0.1253	&	-0.8868     &	Star-forming	&	Star-forming	\\
SWIRE3-J033535.68-284714.2  & 0.123	& -0.8131	&	-0.3639	&	 0.2009	    &	Star-forming	&	Star-forming	\\
SWIRE3-J033523.29-284827.2  & 0.195	& -0.4371	&	-0.1988	&	 0.5581     &	AGN		&	AGN	 	\\
SWIRE3-J033536.42-283309.5  & 0.105	& -0.1468	&	-0.4471	&	 0.0645     &	Composite	&	Star-forming	\\
SWIRE3-J033532.56-283051.1  & 0.077 	& -0.5036	&	-0.3016	&	 0.7621     &	AGN     	&	AGN	 	\\
SWIRE3-J033529.98-283344.4  & 0.225 	& -0.1783	&	-0.3706	&	-0.0500     &	Composite  	&	Star-forming	\\
SWIRE3-J033234.23-293450.8  & 0.529	& -0.1029	&	-0.1649	&	 0.0700     &	Composite	&	LINER		\\
SWIRE3-J033558.74-290742.4  & 0.222	& -0.1365	&	-0.1936	&	 0.0584	    &	Composite	&	LINER		\\
SWIRE3-J033552.63-290535.6  & 0.189     & -0.3475	&	-0.4514	&	 0.0045     &	Composite	&	Star-forming	\\
SWIRE3-J033555.43-290758.9  & 0.190 	& -1.0127	&	-0.4227	&	-0.2384     &	Star-forming	&	Star-forming	\\
SWIRE3-J033551.15-290735.6  & 0.195     & -0.5292	&	-0.3433	&	 0.1401	    &	Star-forming  	&	Star-forming	\\
SWIRE3-J033008.28-285400.3  & 0.217     & -0.7703	&	-0.6176	&	 0.2039     &	Star-forming	&	Star-forming	\\

\hline\hline
\end{tabular}
\end{minipage}
\end{table*}

\begin{figure*}
\centering
\begin{center}
\includegraphics[scale=0.43]{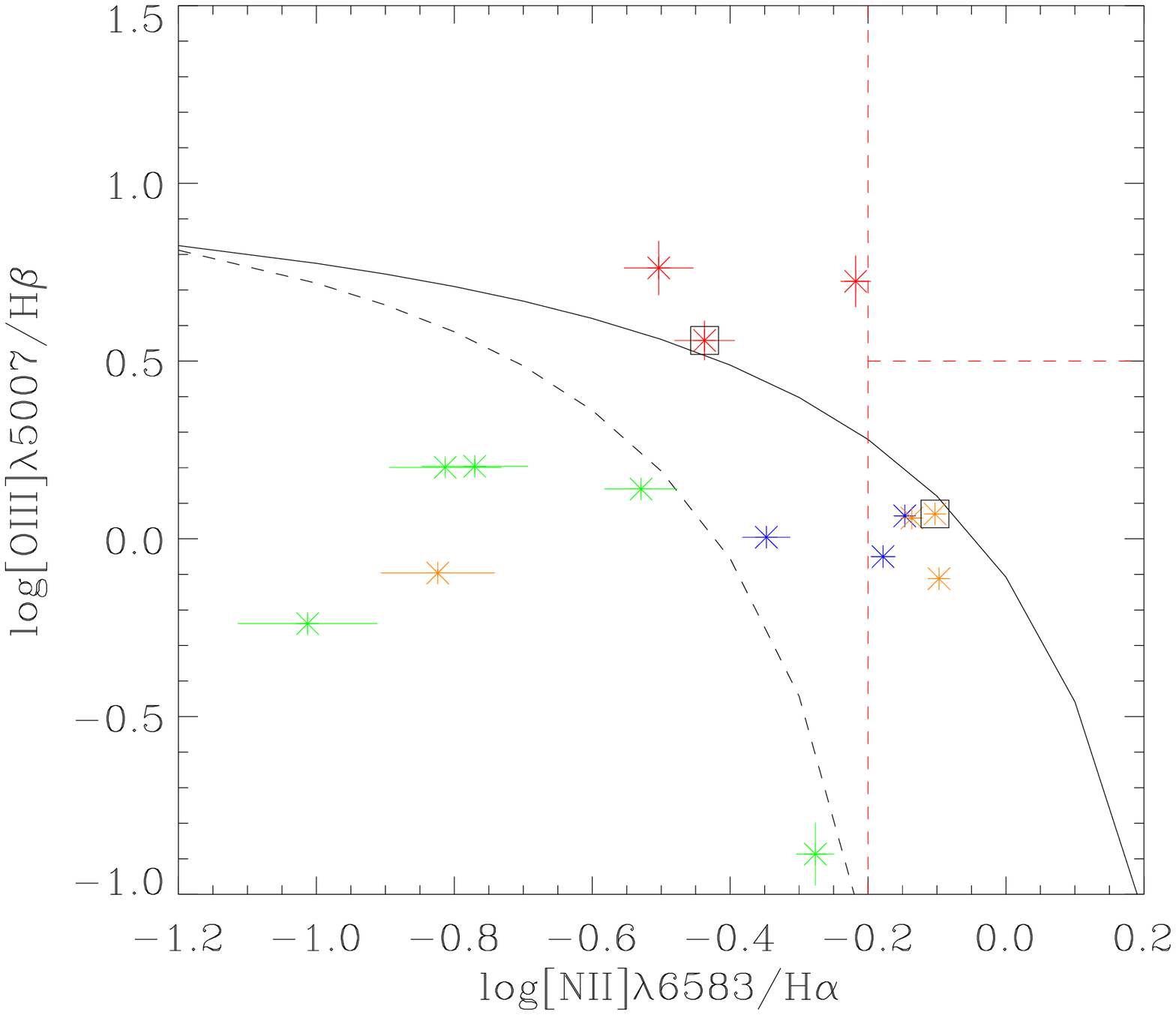}
\includegraphics[scale=0.43]{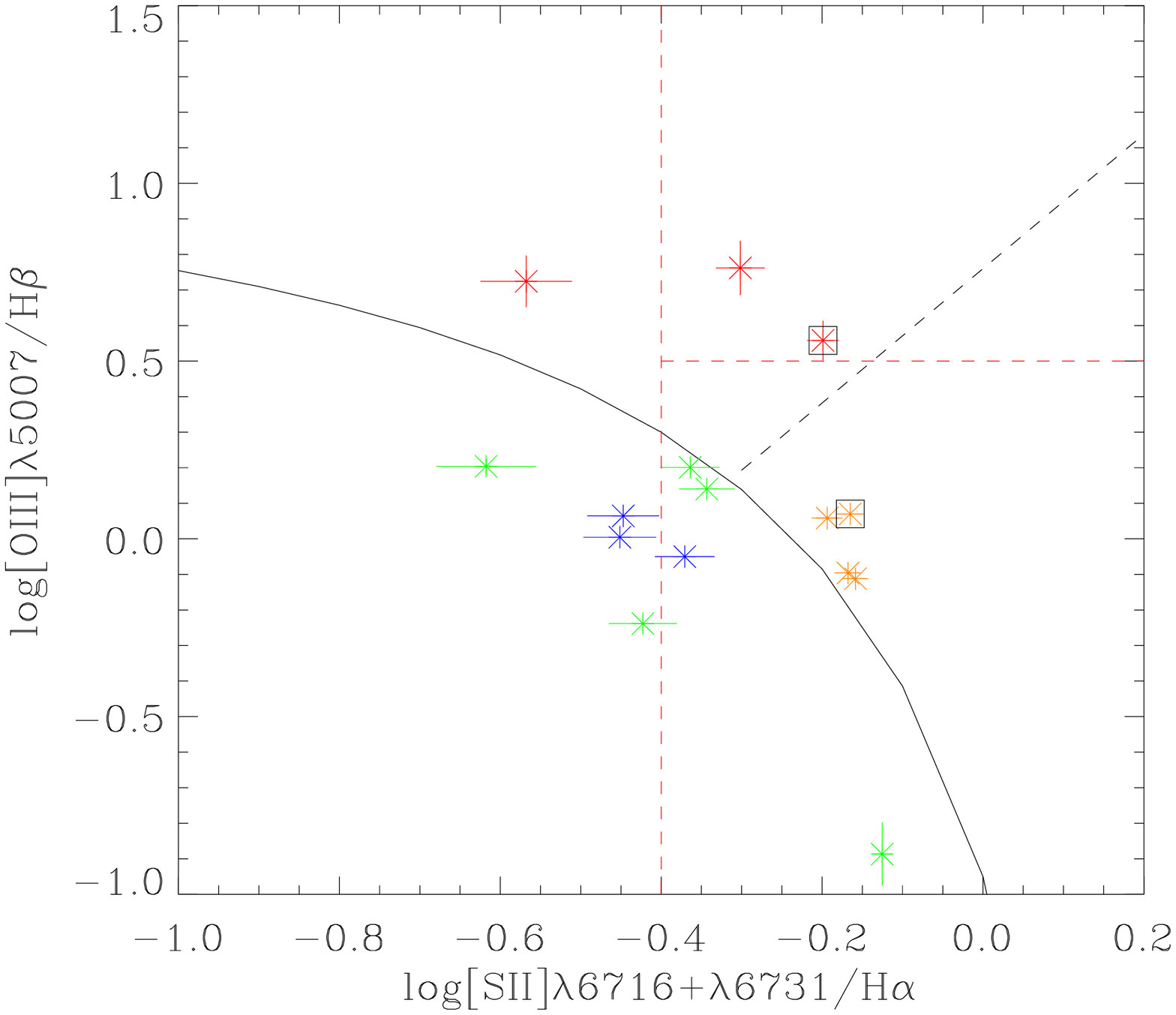}
\end{center}

\caption{\textbf{Left}: The [\textit{NII}]$\lambda$6583/\textit{H$\alpha$} vs [\textit{OIII}]$\lambda$5007/\textit{H$\beta$} diagnostic emission line diagram for our sample of 15 sources with available lines. The black-dashed line is the pure star-forming line (Kauffmann et al. 2003) and the black-solid line is the extreme star-forming line (Kewley et al. 2001).The red-dashed lines are the Seyfert/LINER lines (Ho et al. 1997). Green, blue and red asterisks represent the star-forming, the composite and the Seyfert sources based  on the [\textit{NII}]$\lambda$6583/\textit{H$\alpha$} vs [\textit{OIII}]$\lambda$5007/\textit{H$\beta$} diagnostic. Orange asterisks are LINERs based on the [\textit{SII}]$\lambda$6716+$\lambda$6731/\textit{H$\alpha$} vs [\textit{OIII}]$\lambda$5007/\textit{H$\beta$} diagnostic. Open black squares are U/LIRGs of our sample with available lines. \textbf{Right}: The [\textit{SII}]$\lambda$6716+$\lambda$6731/\textit{H$\alpha$} vs [\textit{OIII}]$\lambda$5007/\textit{H$\beta$} diagnostic emission line diagram for the sample of 15 sources. The black-solid line is the AGN/Starburst line (Baldwin et al. 1981) and the black-dashed line is the Seyfert/LINER line (Kewley et al. 2006). We keep the same colors for the sources.}
\end{figure*}

\section{EMISSION LINE DIAGNOSTICS}

A suite of emission line diagnostic diagrams has been used extensively in order to classify the dominant energy sources in emission-line galaxies (Baldwin et al. 1981). These diagrams are based on various optical line ratios. We have used the following three: [\textit{OIII}]/\textit{H$\beta$}, [\textit{NII}]/\textit{H$\alpha$} and [\textit{SII}]/\textit{H$\alpha$}. The reason behind this selection is that these ratios are not affected by reddening and spectrophotometric effects even in the case of non-flux calibrated spectra.\\
For the [\textit{SII}]/\textit{H$\alpha$} versus [\textit{OIII}]/\textit{H$\beta$} and [\textit{NII}]/\textit{H$\alpha$}
versus [\textit{OIII}]/\textit{H$\beta$} diagrams the theoretical work of (Kewley et al. 2001) provided a maximum starburst line and clear division between AGN and SF galaxies with all galaxies lying above this line to be dominated by an AGN. The work of Kauffmann et al. (2003) provided a cleaner delineation between AGN and SF galaxies due to the large sample of SDSS galaxies. The result was an empirical relation dividing pure star-forming galaxies from Seyfert-HII composite objects whose spectra contain significant contributions from both AGN and star formation. In the [\textit{SII}]/\textit{H$\alpha$} versus [\textit{OIII}]/\textit{H$\beta$} diagram, the AGNs lie on two branches: on the upper one Seyfert galaxies while LINERs lie on the lower one. The division between the two AGN branches takes effect via the Seyfert-LINER classification line (Kewley et al. 2006).\\
In order to recreate the BPT diagrams for our sample, we have plotted all narrow-line sources with detections in all 5 lines: [\textit{OIII}], \textit{H$\beta$},
[\textit{NII}], \textit{H$\alpha$}, [\textit{SII}], a total of 15 sources. The flux in each emission line is measured using the NOAO IRAF SPLOT task, fitting a Gaussian function to each emission line. Diagnostic line ratios for [\textit{NII}]$\lambda$6583/\textit{H$\alpha$}, [\textit{SII}]/\textit{H$\alpha$} and [\textit{OIII}]$\lambda$6300/\textit{H$\beta$} are plotted in Figure 5 and the data summarized in Table 3.\\
Figure 5 (left) shows the [\textit{NII}]$\lambda$6583/\textit{H$\alpha$} vs [\textit{OIII}]/\textit{H$\beta$} diagnostic for our sample 
of 15 sources. The extreme (Kewley et al. 2001) and pure (Kauffmann et al. 2003) star-forming lines are shown as black-dashed and black-solid lines 
respectively. Galaxies that lie below the pure star-forming lines are HII region-like galaxies. Galaxies that lie in between the two 
classification lines, extreme and pure star-forming lines, are known as composite objects being on the AGN-HII mixing sequence.
 Composite galaxies' spectra are likely to contain significant contributions from a metal-rich stellar population plus an AGN. 
Everything that lies above the solid line (Kewley et al. 2001) is classed as AGN. Ho et al. (1997) defined a new classification scheme which is 
represented by the red-dashed lines. Adding these lines to the previous two classification lines we separate the classification scheme 
in four parts. As HII-region are classified the sources which lie below the Kewley line and left from the red-dashed line. Seyfert 
sources lie at the upper left box and the LINERs at the low right box. Figure 5 (right) shows the [\textit{SII}]/\textit{H$\alpha$} 
vs [\textit{OIII}]/\textit{H$\beta$} diagnostic emission line diagram for the sample 15 sources. The black-solid line (Baldwin et al. 1981) 
provides an upper limit to the star-forming sources on this diagram. The black-dashed line (Kewley et al. 2006) provides a division between 
Seyfert and LINER sources. The red-dashed lines represent the Ho et al. (1997) classification line in [\textit{NII}]/\textit{H$\alpha$} vs 
[\textit{OIII}]/\textit{H$\beta$} diagram.\\
Based on both diagnostic diagrams we found 6 pure star-forming galaxies of which one appears as LINER on the [\textit{SII}]/\textit{H$\alpha$} vs [\textit{OIII}]/\textit{H$\beta$} diagram, 3 Seyfert galaxies and 6 composite objects of which 3 appear 
as star-forming objects and 3 as LINERs on the [\textit{SII}]/\textit{H$\alpha$} vs [\textit{OIII}]/\textit{H$\beta$} diagram. 
The spectra of these objects are available in Figure 10.

\begin{figure}
\centering
\begin{center}
\includegraphics[scale=0.43]{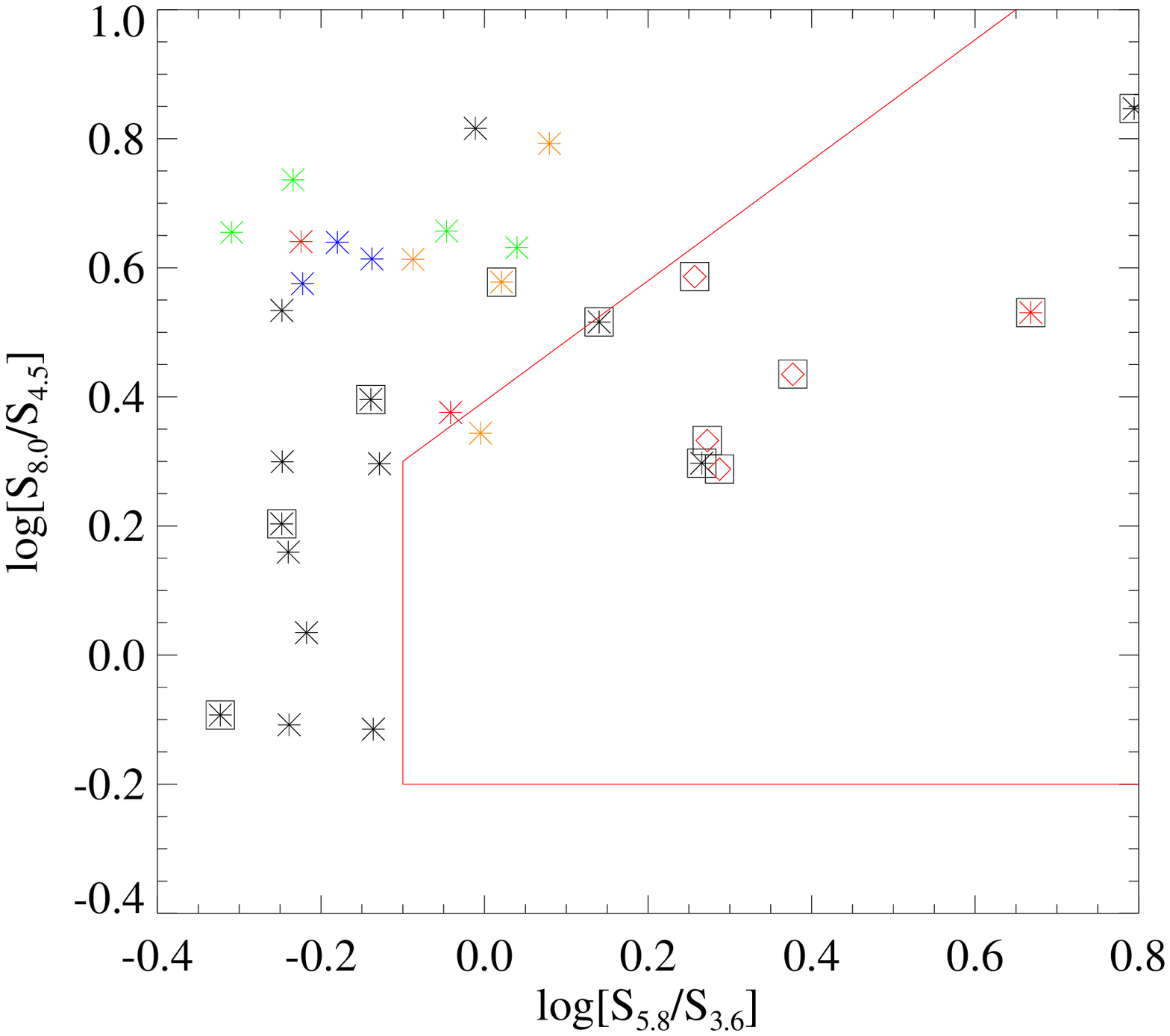}

\end{center}
\caption{IRAC colour-colour plot (Lacy et al. 2004) of all 34 sources of our sample with detections in all IRAC bands. Green, blue, orange and red colors represent the star-forming, composite, LINER and Seyfert narrow line emission sources based on [\textit{NII}]/\textit{H$\alpha$} vs [\textit{OIII}]/\textit{H$\beta$} and [\textit{SII}]/\textit{H$\alpha$} vs 
[\textit{OIII}]/\textit{H$\beta$} diagrams. Red diamonds are broad line QSOs. Black asterisks represent all the rest sources. Red solid line is the AGN area as defined by Lacy et al. (2004). Open black squares represent the H/U/LIRGs of our sample.}
\end{figure}

\begin{figure}
\centering
\begin{center}
\includegraphics[scale=0.43]{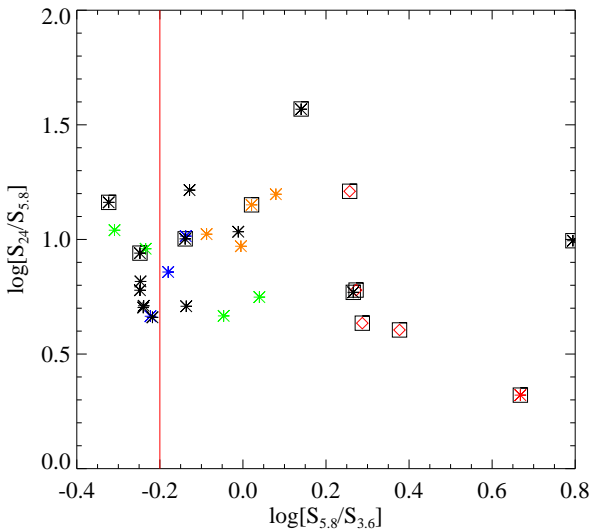}

\end{center}
\caption{IRAC-MIPS colour-colour plot of 31 sources of our sample with detections in 3.6$\mu$m, 5.8$\mu$m and 24$\mu$m bands. 
Red solid line distinguishes between AGN and star-forming galaxies (Lacy et al. 2004). Colors and symbols are same as Figure 6.}
\end{figure}
\section{MIR COLORS}

For the Lacy et al. (2004) diagnostic we require sources to have detections in all four IRAC bands, 3.6$\mu$m, 4.5$\mu$m, 5.8$\mu$m and 
8$\mu$m, a total of 29 sources.  For the remaining 5 we have used SWIRE limits for each band. Sources that lie within the Lacy wedge 
are those expected to be AGN dominated from the mid-infrared colors.\\
Figure 6 shows the IRAC colour-colour plot combined with information obtained from the spectroscopic diagnostics. In our sample we 
detect 4 broad line objects which are fitted with a QSO optical template and all of them lie within the AGN wedge (red diamonds). 
In the case of the narrow emission lines, all the star-forming sources (green asterisks) lie outside the AGN region and specifically on the top-left of 
the diagram. In the same region with star-forming sources lie the LINER sources at a rate of 2/3 (orange asterisks) and all the composite sources (blue asterisks). 
Only 1 of the 3 narrow-line AGNs we found in our sample (based on BPT diagrams) lies within the AGN wedge. All the H/ULIRGs of our sample lie within 
the AGN region in contrast with LIRG sources which lie outside the AGN wedge at a rate of 5/6. 
In Figure 7 we plot an IRAC-MIPS color-color diagram using the sources with detection in IRAC's 3.6$\mu$m, 5.8$\mu$m bands and MIPS's 
24$\mu$m. This requirement limits the number of our sample to 31. Even if the larger wavelength difference between 8 and 24$\mu$m makes 
the interpretation of the MIPS/IRAC color-color plots more complicated, all the QSO fitted sources lie within the AGN region (right of 
the red line). The rate of LIRG sources that lie with in this region has been increased at 4/6. Figure 8 shows the IRAC color-color 
diagnostic diagram from Stern et al. (2005). Again, all the broad-line objects lie within the AGN region and 1/3 Type-II Seyferts. 
Figure 9 shows the SEDs of the 13 sources with z$>$0.5. All of them are fitted with the standard infrared templates of Rowan-Robinson et al. (2008).

\begin{figure}
\centering
\begin{center}
\includegraphics[scale=0.43]{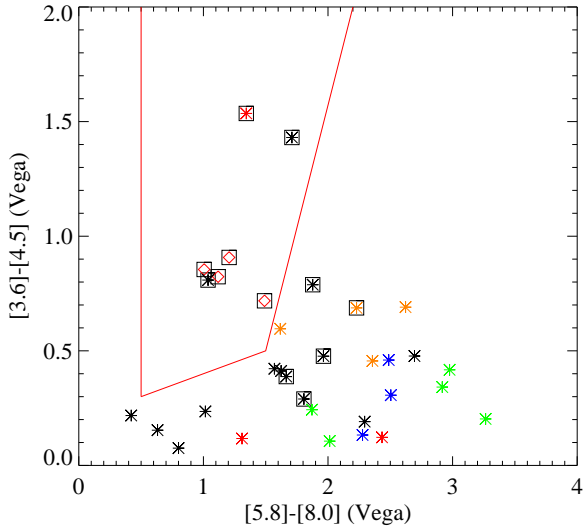}

\end{center}
\caption{IRAC colour-colour plot (Stern et al. 2005) of 34 sources of our sample with detections in 3.6$\mu$m, 4.5$\mu$m, 5.8$\mu$m and 
8.0$\mu$m bands. Red solid lines follow empirical criteria to separate AGNs from other sources (Stern et al. 2005). 
Colors and symbols are same as Figure 6.}
\end{figure}

\begin{figure}
\centering
\begin{center}
\includegraphics[scale=0.43]{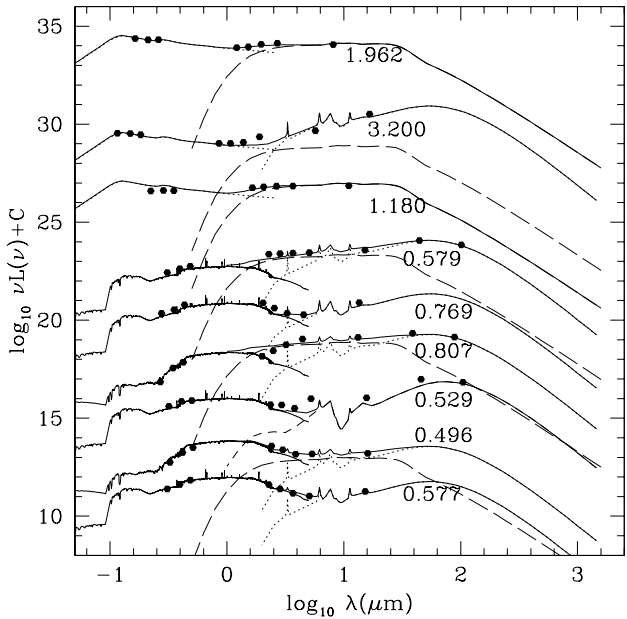}

\end{center}
\caption{SEDs for the 9 galaxies with z$>$0.5. Two are fitted with an M82 starburst, one with an Arp220 starburst, two with an AGN 
dust torus and four are composite M82 starburst+AGN dust torus.  All three with QSO optical templates have AGN dust tori 
(the long-dashed curves).}
\end{figure}

\section{Summary}

We present the optical spectra of 34 sources  within SWIRE-CDFS, observed using EFOSC-2 in MOS mode on the ESO 3.6m telescope. We have 
compared the observed spectroscopic redshifts values to those estimated using our photoz methods and the agreement is very good 
especially in the case of galaxies. With respect to QSOs' photoz agreement and taking into consideration the problems photoz methods 
face when dealing with quasars, our sample is too small to extract safe results. Among our sources we found 6 LIRGs, 5 ULIRGs and 1 
HLIRG. All H/ULIRGs are broad line objects at z$>$1.0, in excellent agreement with our SED template fitting method which fitted all of 
them with a QSO optical template. All the broad line objects in our sample are fitted with a QSO optical template and show evidence of a 
strong dust torus component in the MIR in excellent agreement with MIR color-color plots. In the case of narrow emission line objects, 
based on emission diagnostic diagrams for the 15 sources with the necessary lines, we found 6 pure star-forming galaxies, 3 Seyfert 
galaxies, 6 composite objects of which 3 appear as star-forming objects and 3 as LINERs based on the [\textit{SII}]/\textit{H$\alpha$} vs 
[\textit{OIII}]/\textit{H$\beta$} diagram.

\begin{figure*}
\centering
\begin{center}

\includegraphics[scale=0.55]{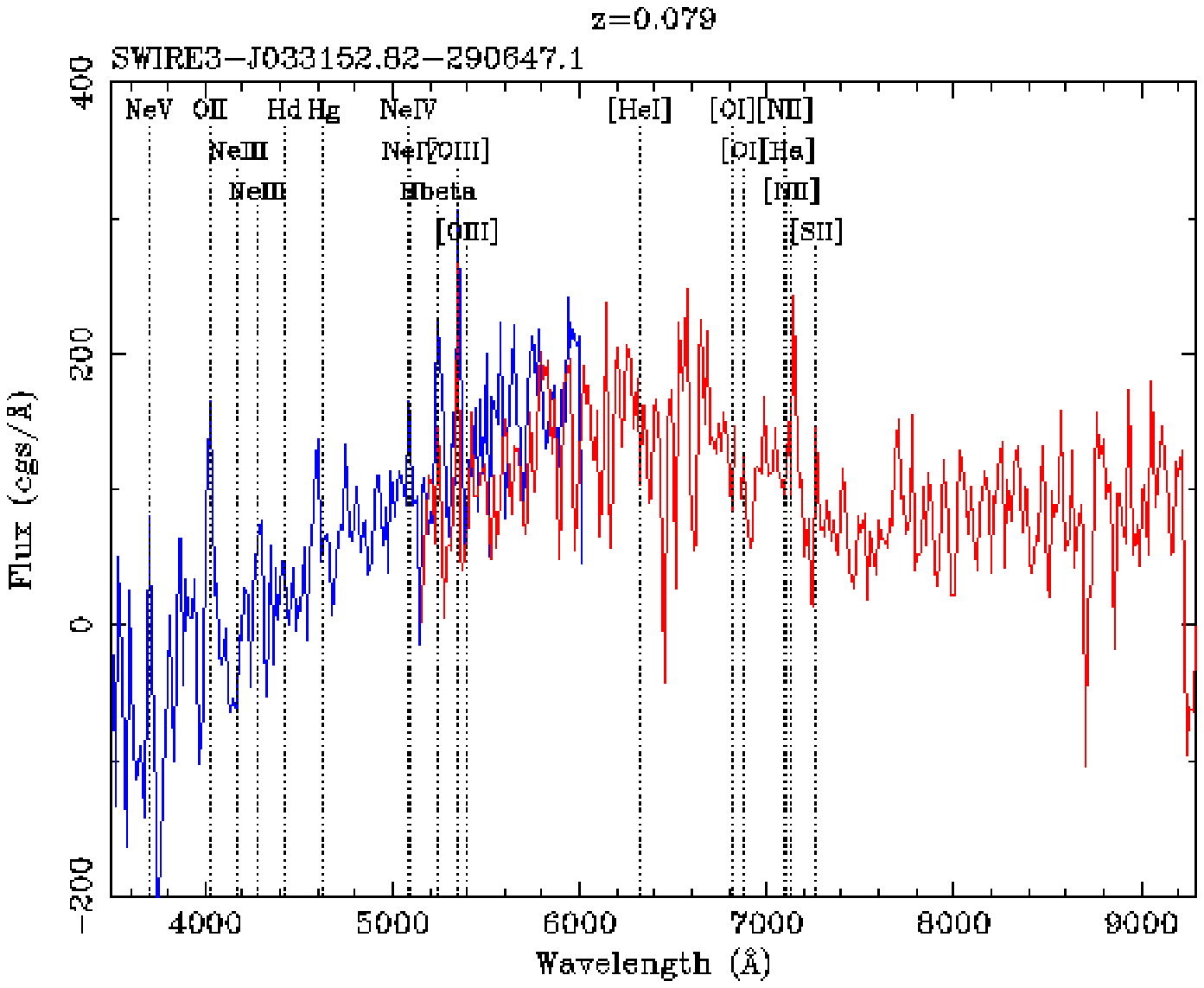}
\includegraphics[scale=0.55]{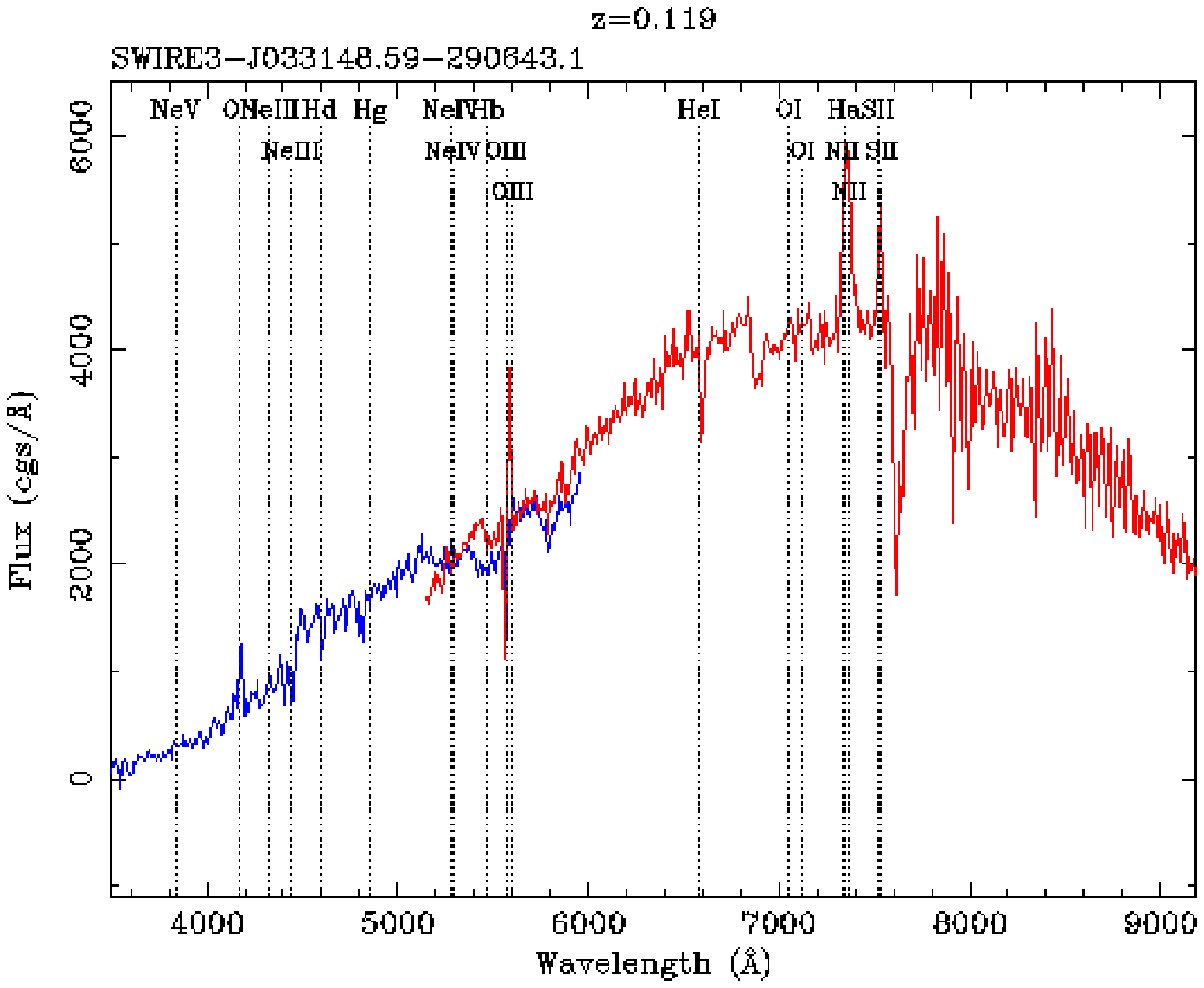}

\vspace{5pt}

\includegraphics[scale=0.55]{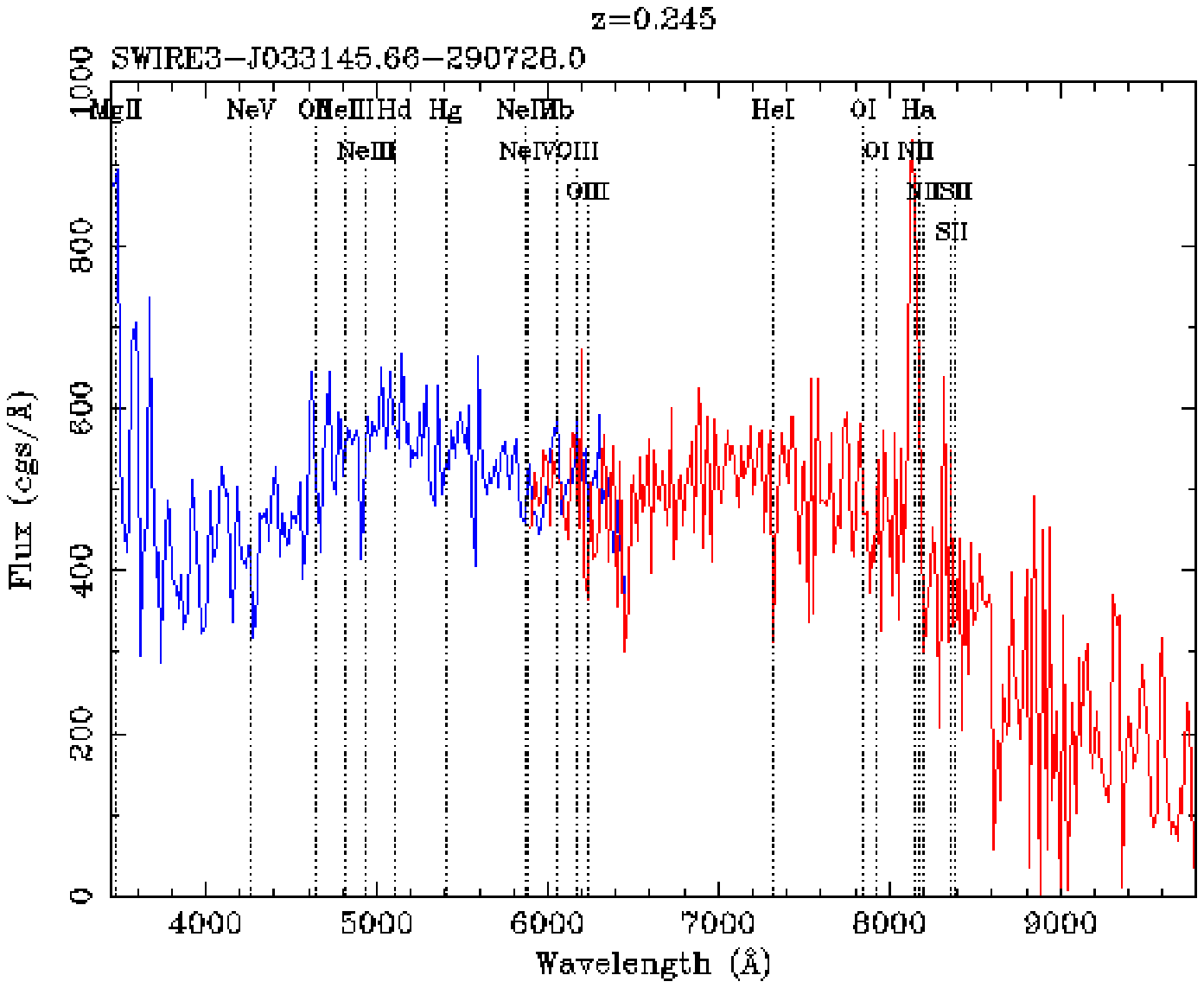}
\includegraphics[scale=0.55]{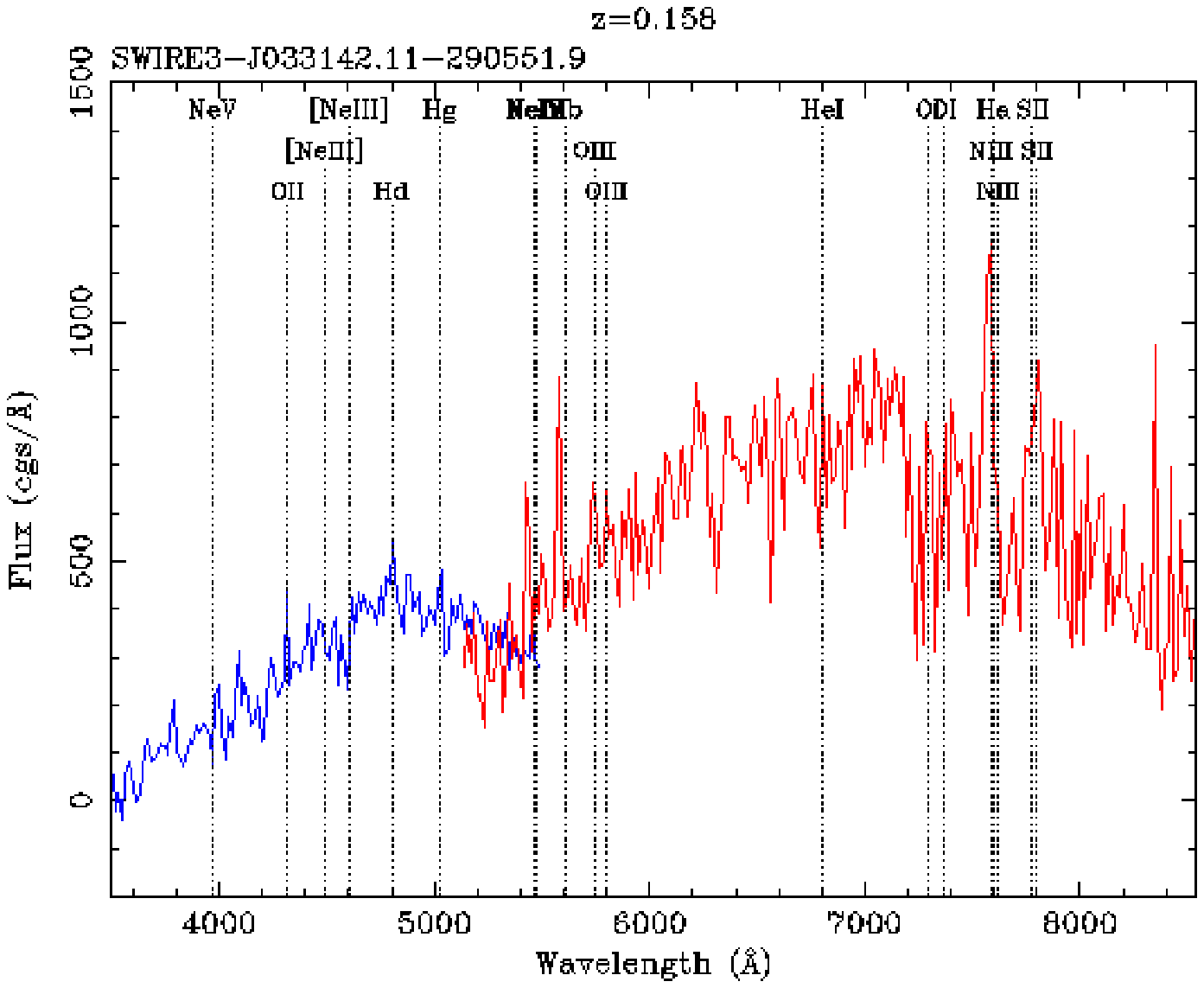}

\vspace{5pt}

\includegraphics[scale=0.55]{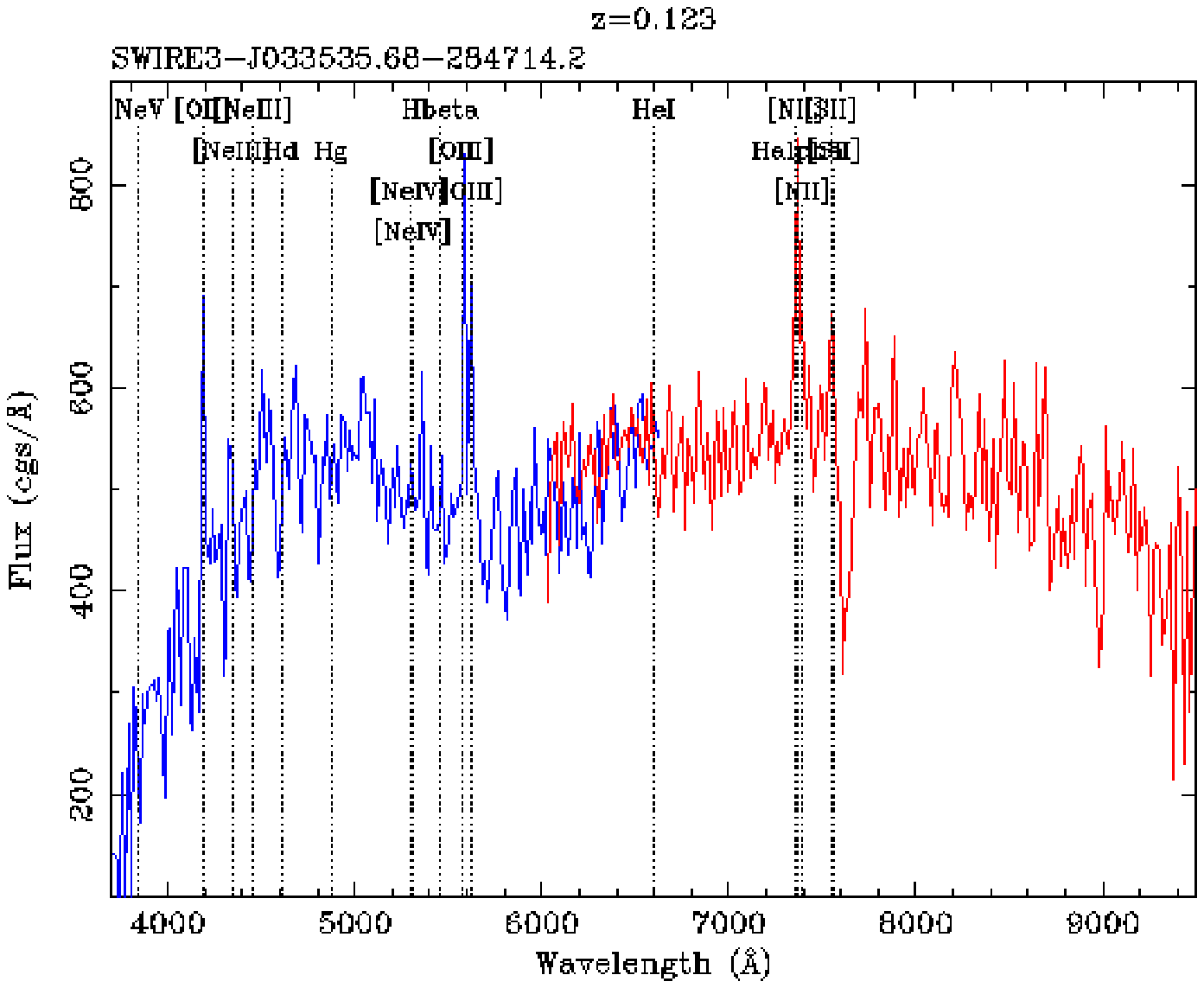}
\includegraphics[scale=0.55]{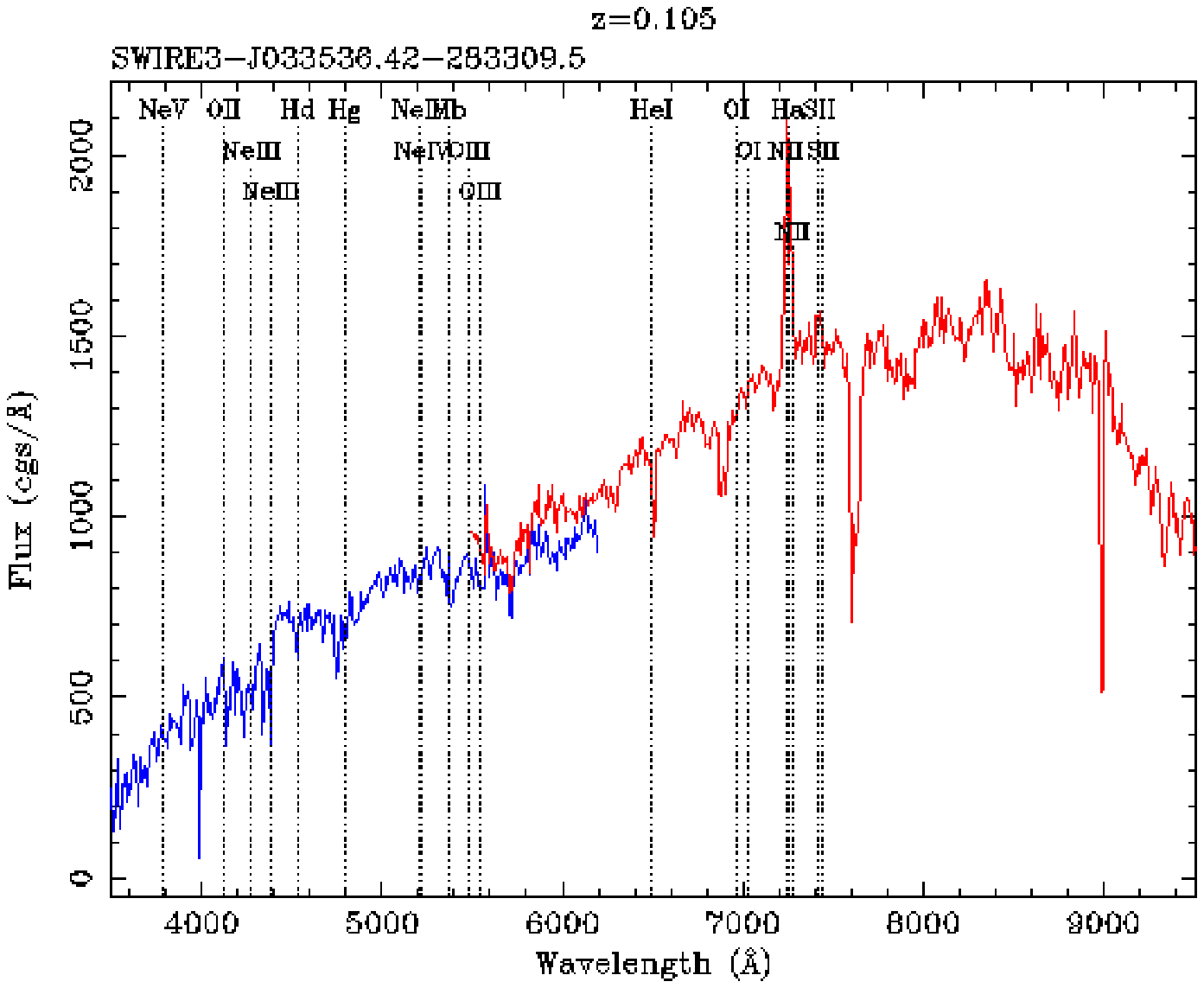}

\end{center}
\end{figure*}

\begin{figure*}
\centering
\begin{center}

\includegraphics[scale=0.55]{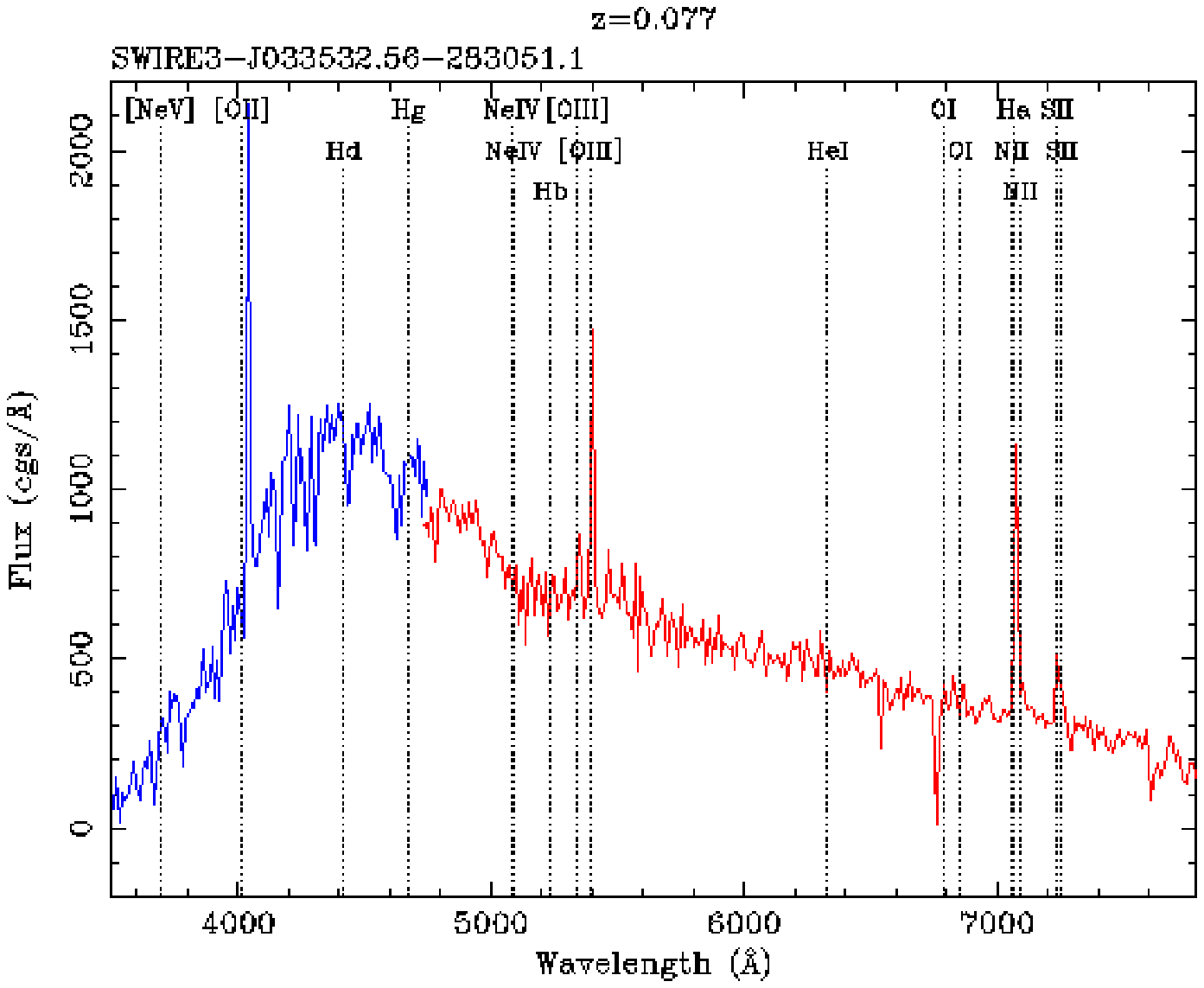}
\includegraphics[scale=0.55]{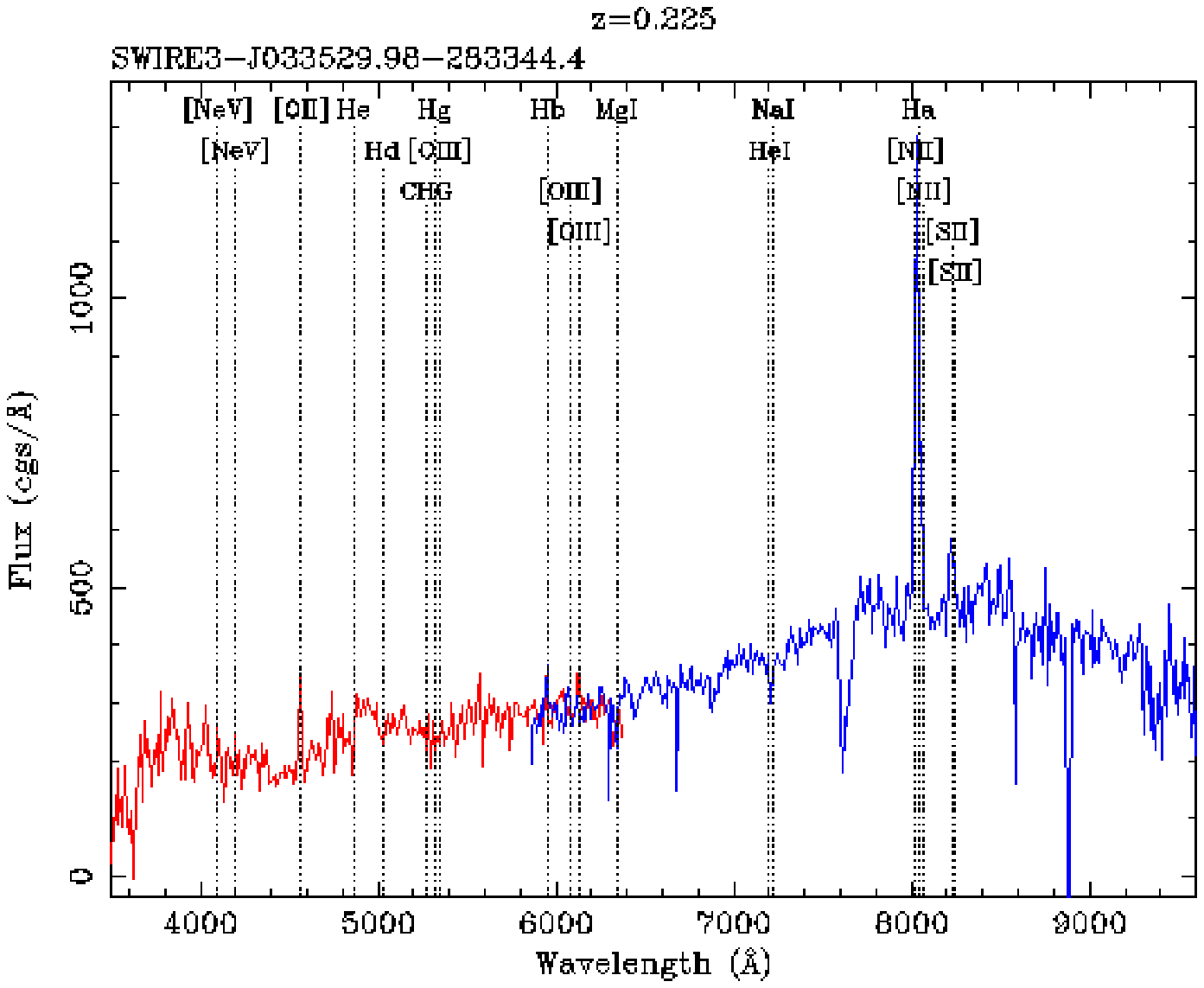}

\vspace{5pt}

\includegraphics[scale=0.55]{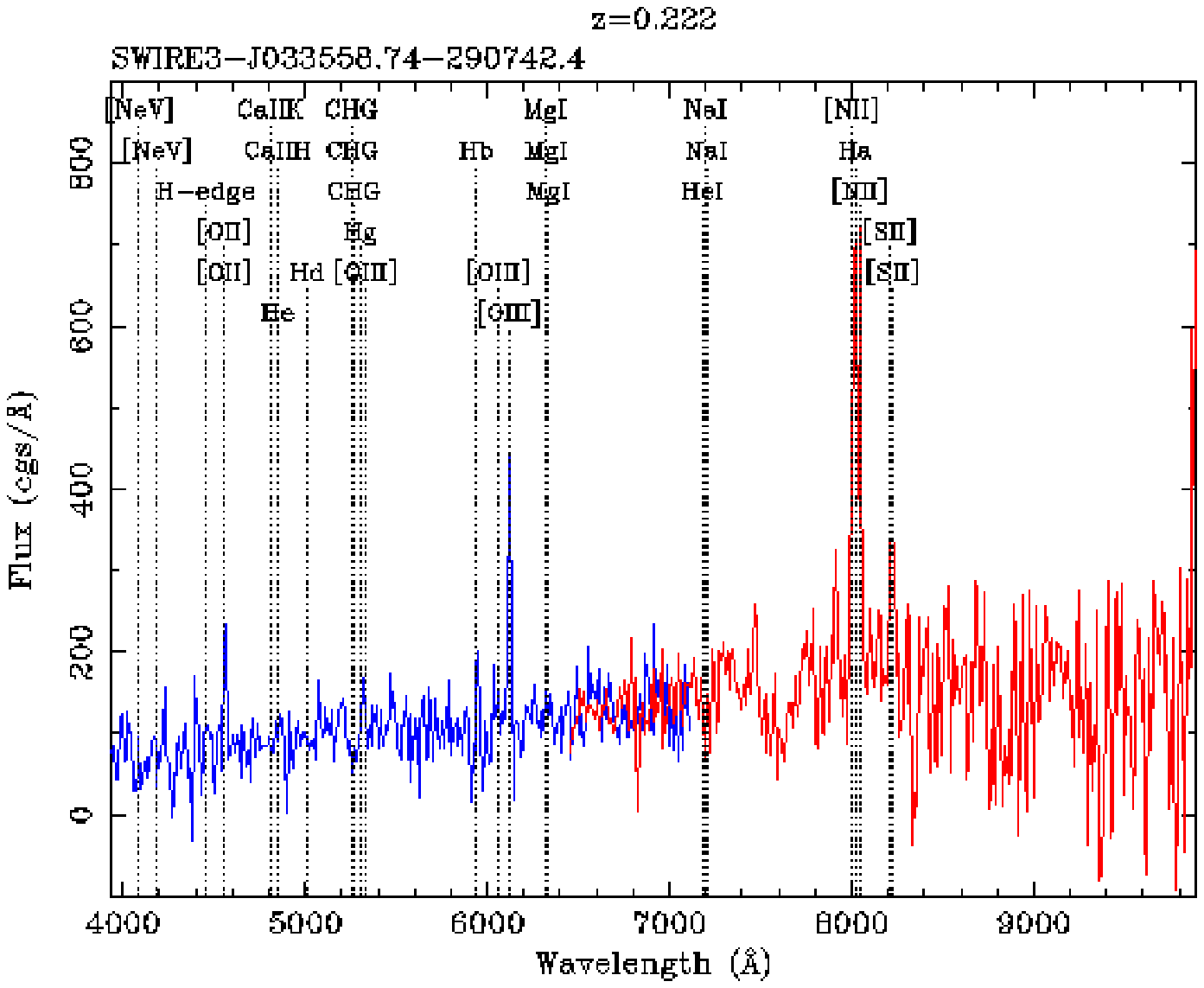}
\includegraphics[scale=0.55]{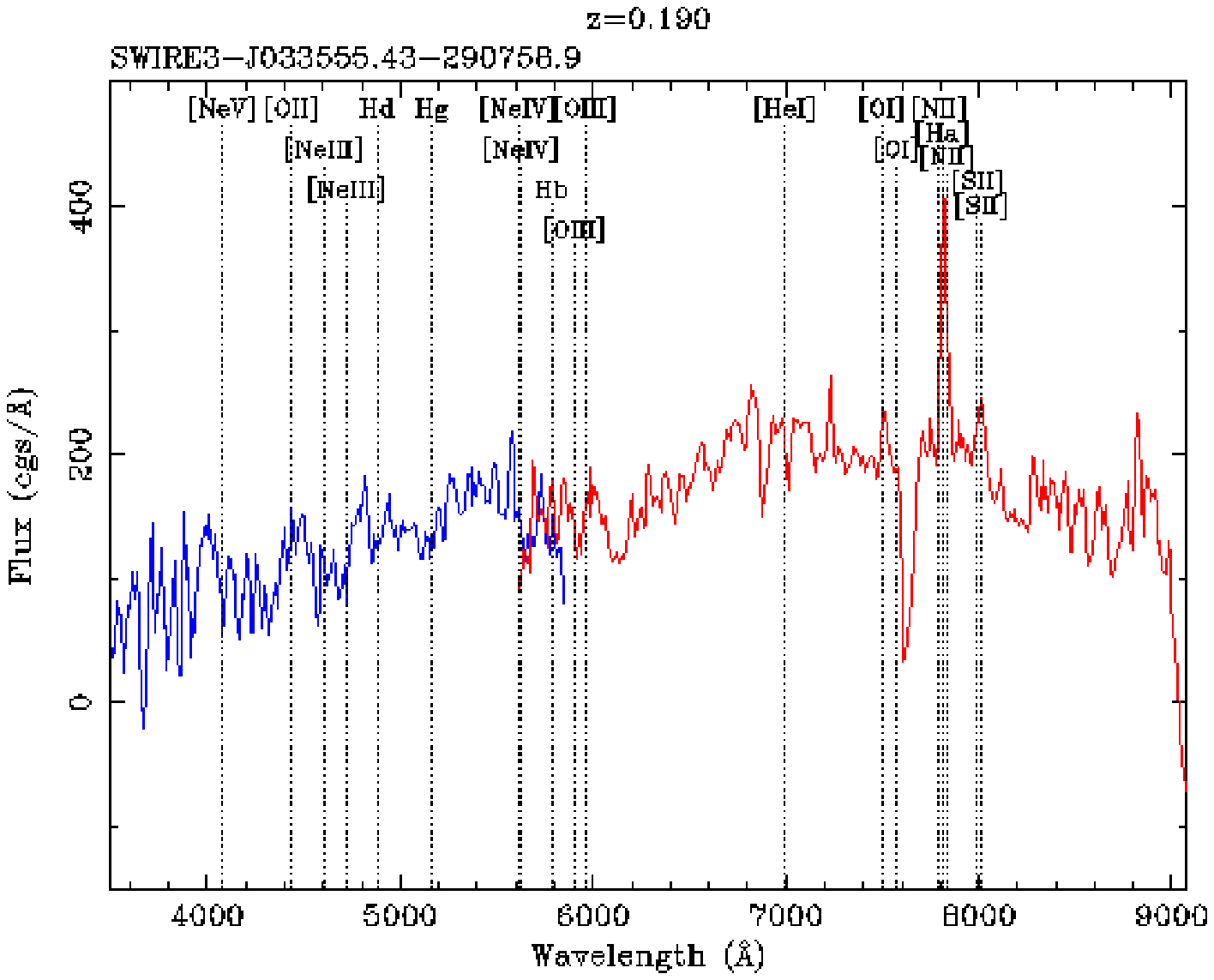}

\vspace{5pt}

\includegraphics[scale=0.55]{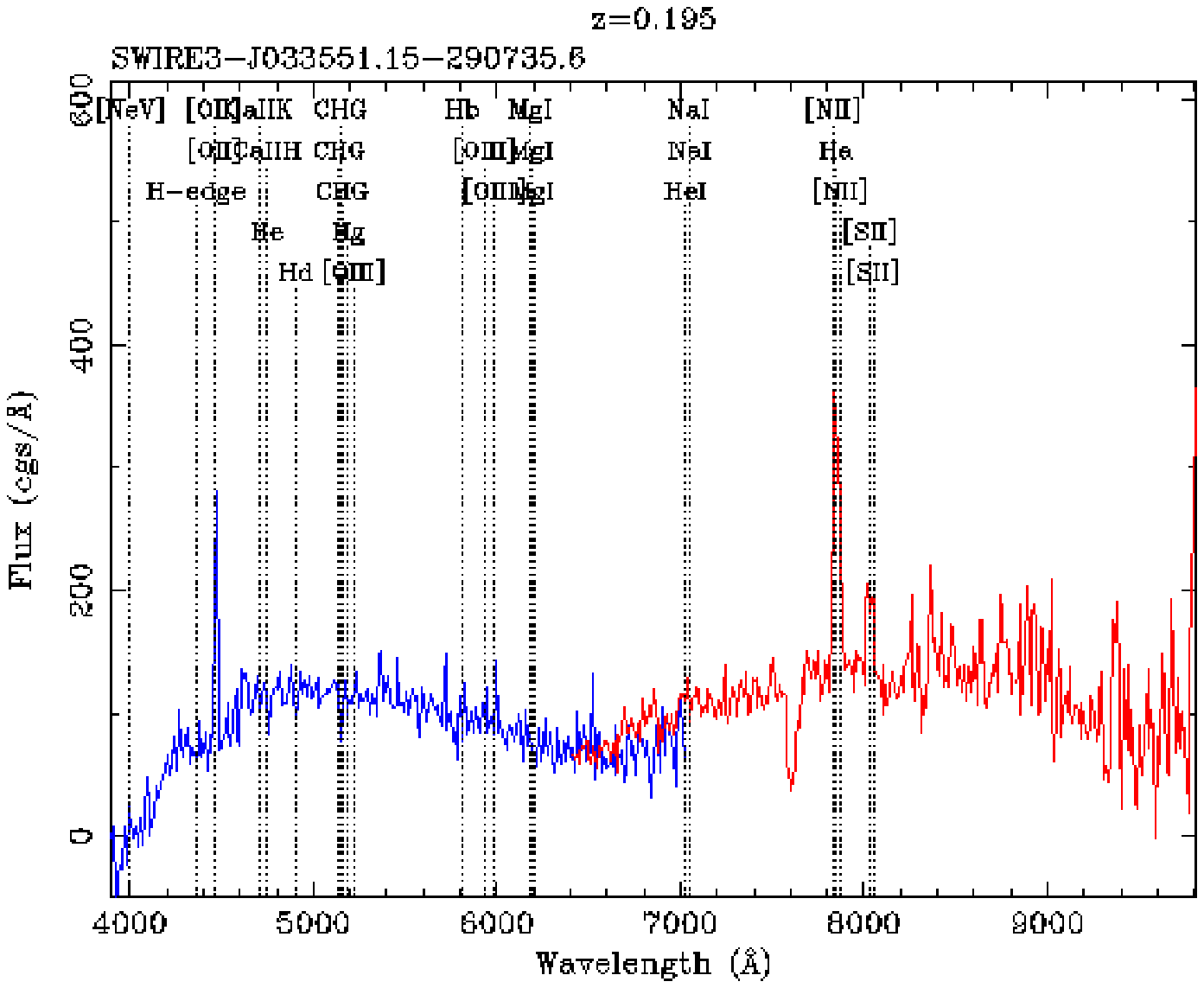}
\includegraphics[scale=0.55]{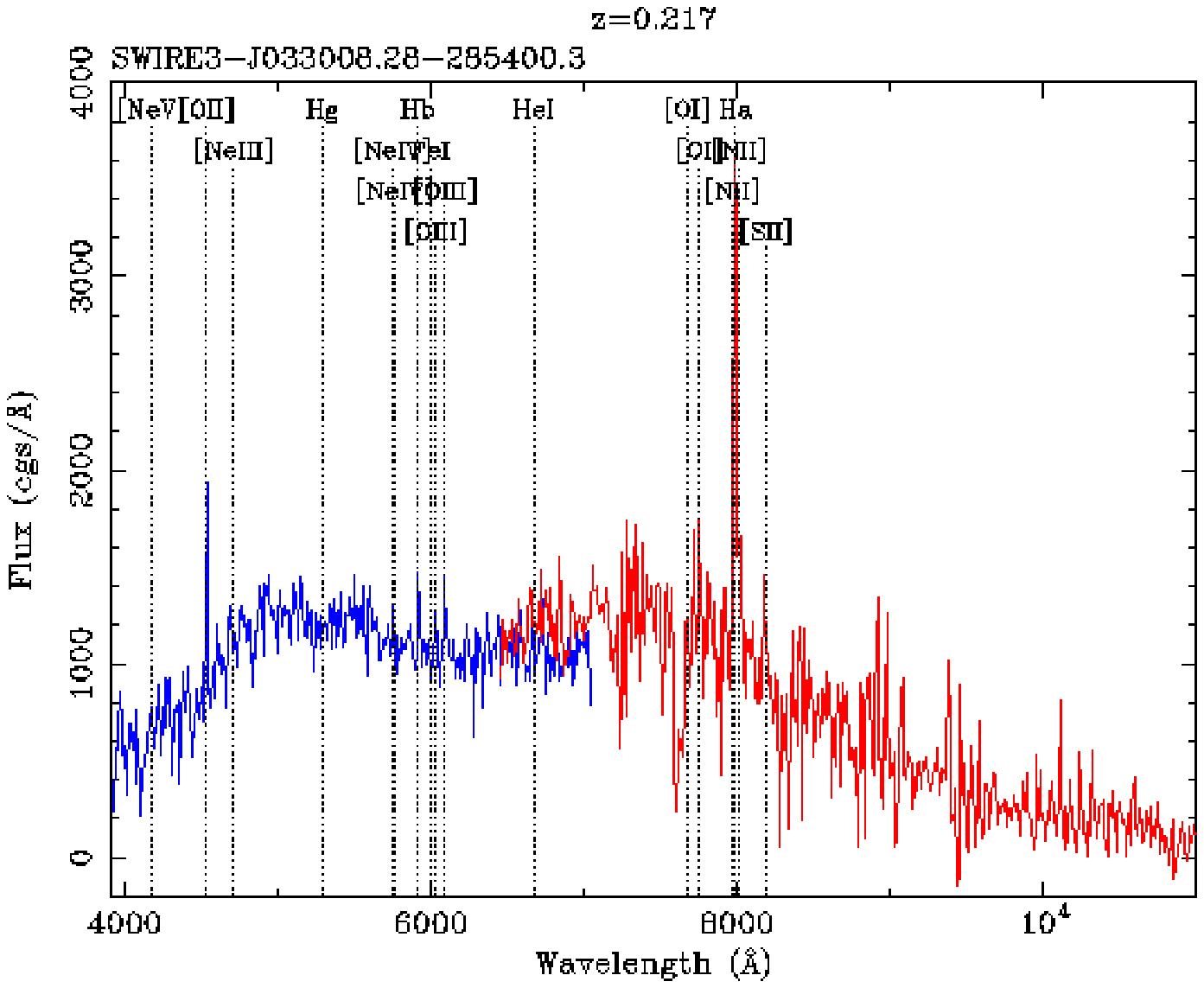}

\end{center}
\end{figure*}

\begin{figure*}
\centering
\begin{center}

\includegraphics[scale=0.55]{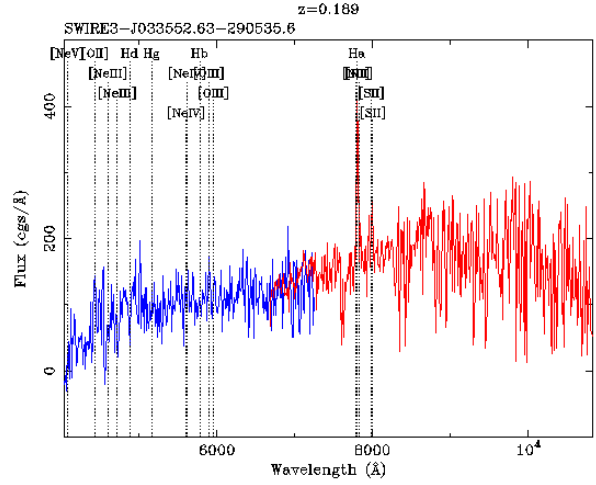}

\end{center}
\caption{A sample of spectra with available [\textit{SII}], \textit{H$\alpha$}, [\textit{OIII}], \textit{H$\beta$}, [\textit{NII}] 
lines, used to estimate line ratios. The redshift of these sources is from 0.077 to 0.529. The 2 U/LIRGs of this sample are plotted 
in Figures 3 and 4.}
\end{figure*}

\begin{figure*}
\centering
\begin{center}

\includegraphics[scale=0.55]{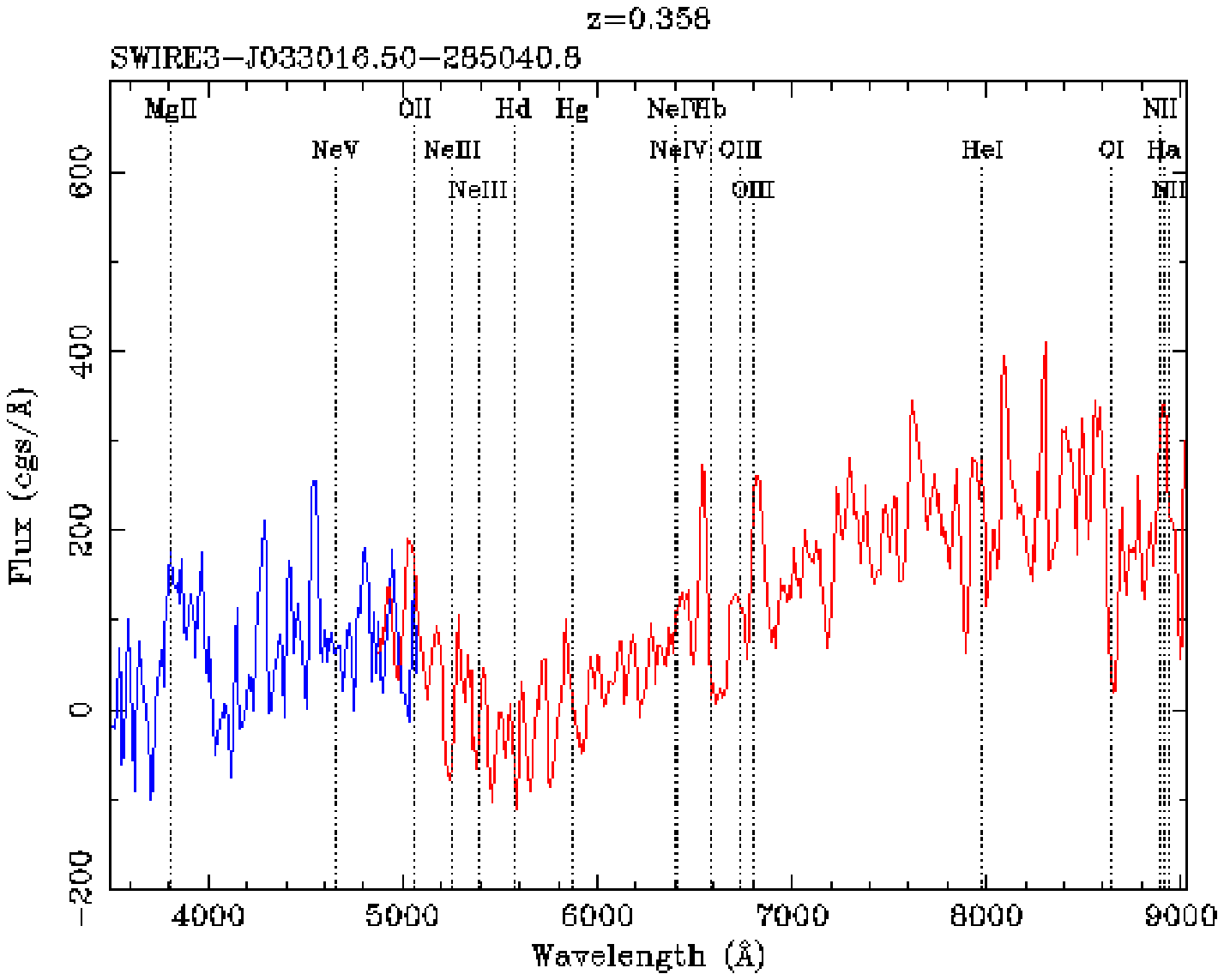}
\includegraphics[scale=0.55]{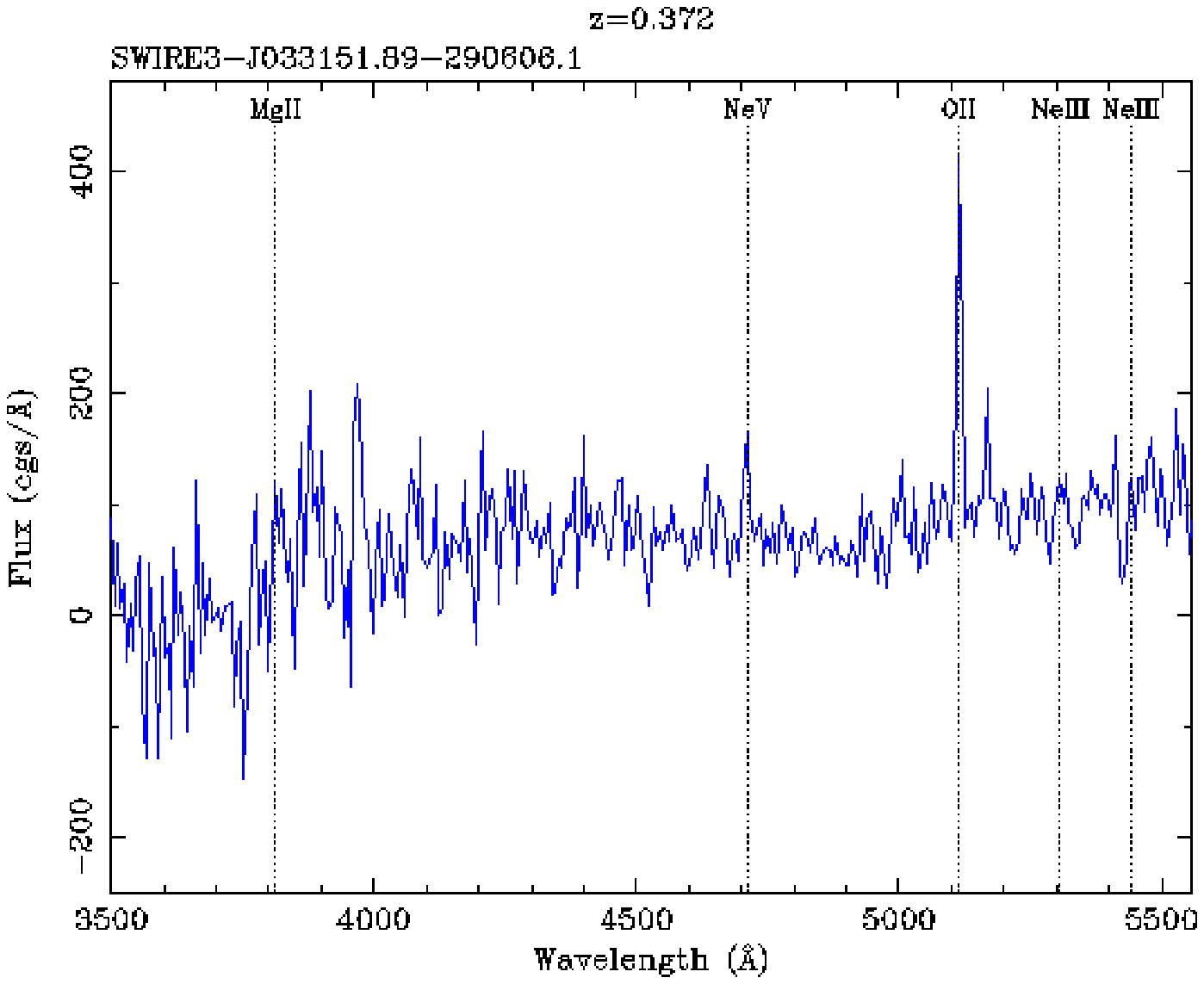}

\vspace{5pt}
\includegraphics[scale=0.55]{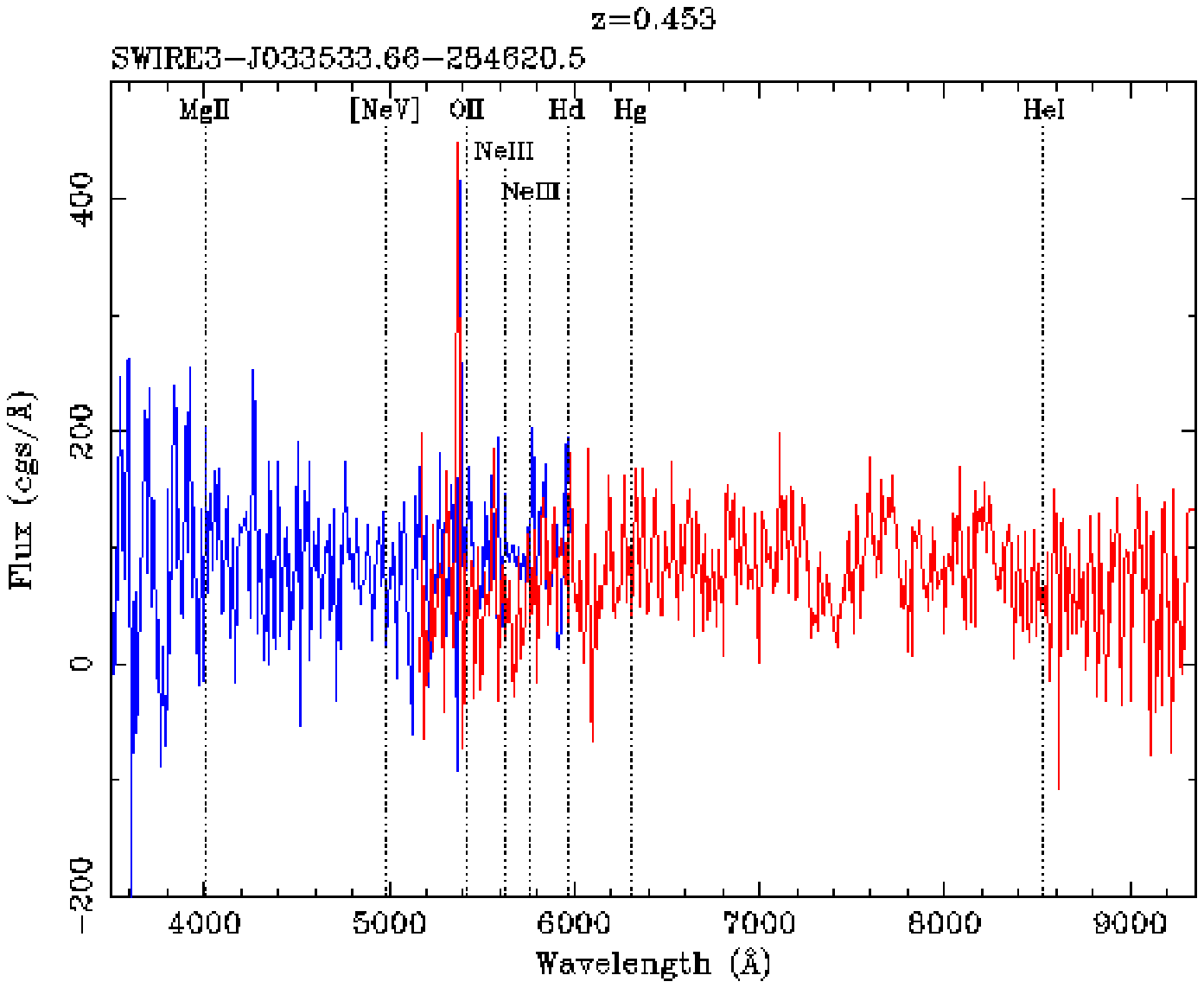}
\includegraphics[scale=0.4]{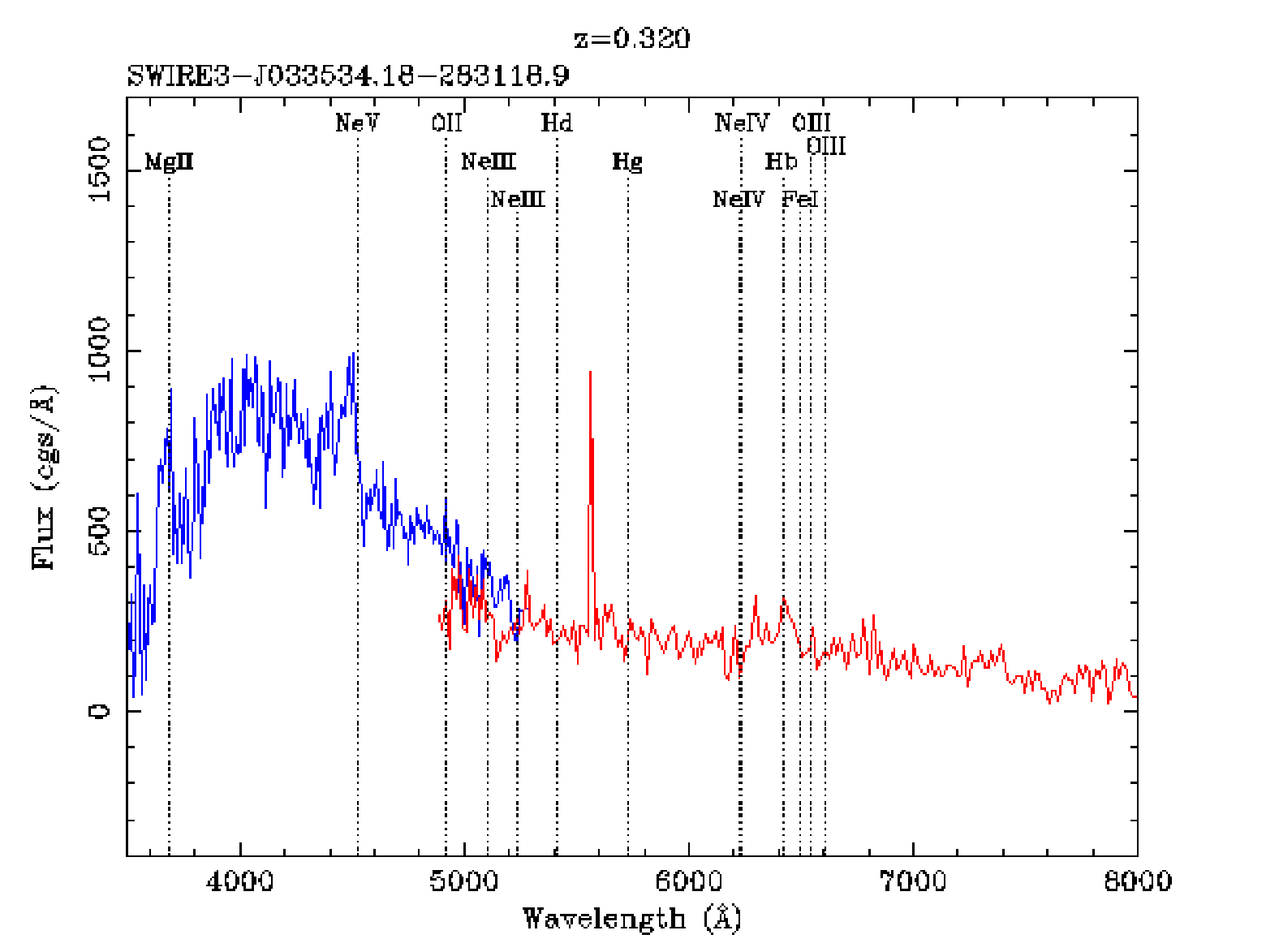}

\end{center}
\end{figure*}

\begin{figure*}
\centering
\begin{center}

\includegraphics[scale=0.55]{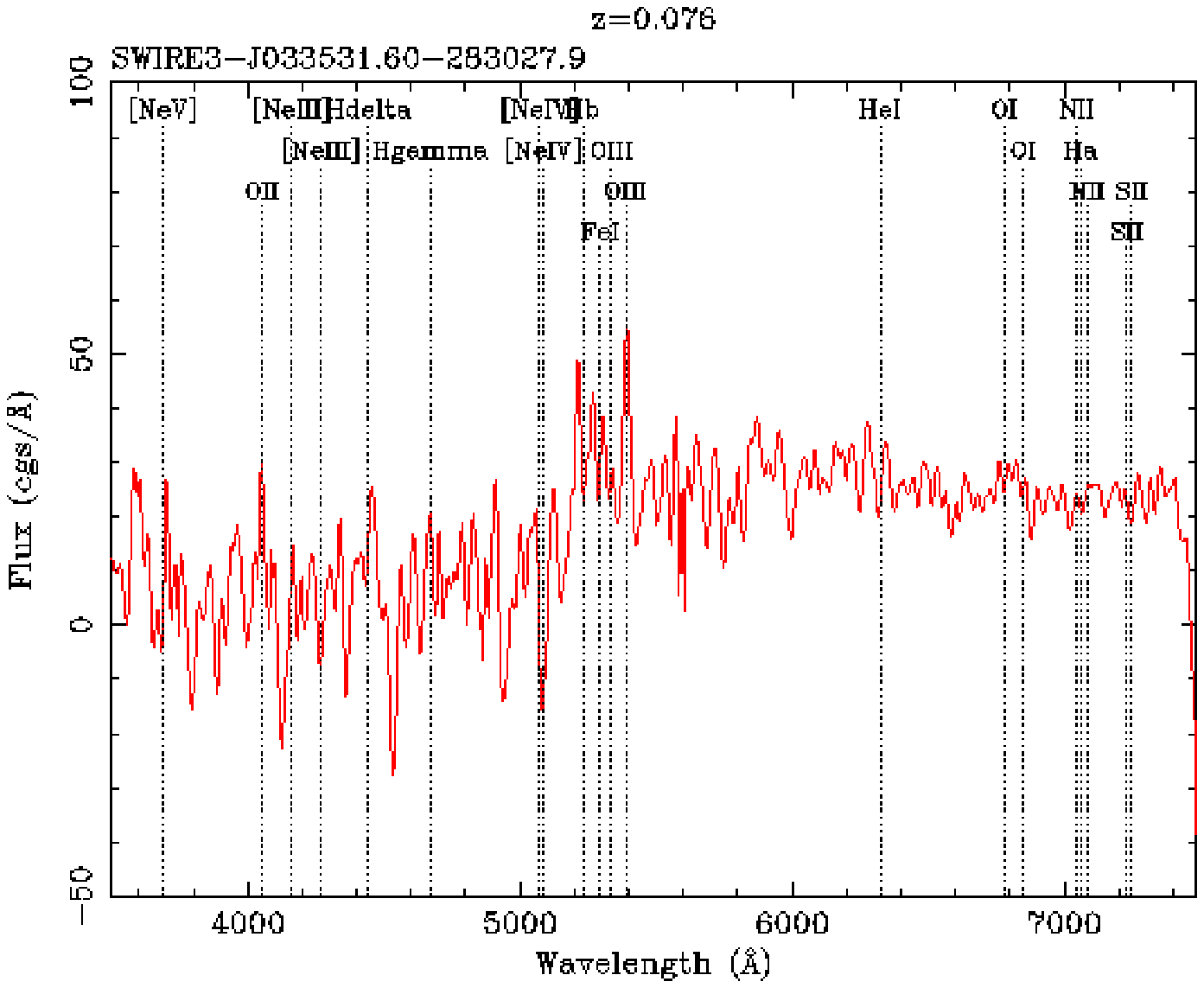}
\includegraphics[scale=0.55]{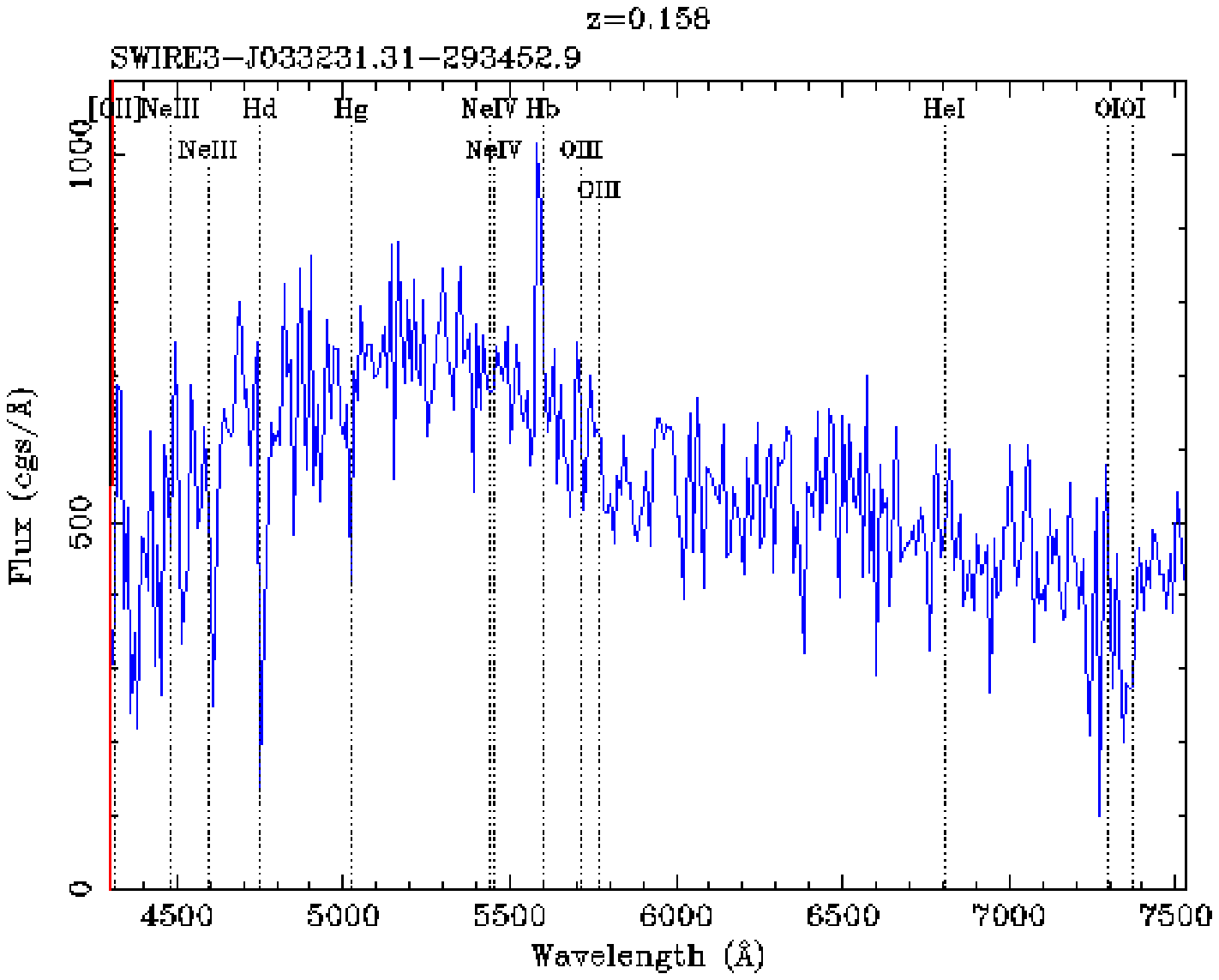}
\vspace{5pt}

\includegraphics[scale=0.55]{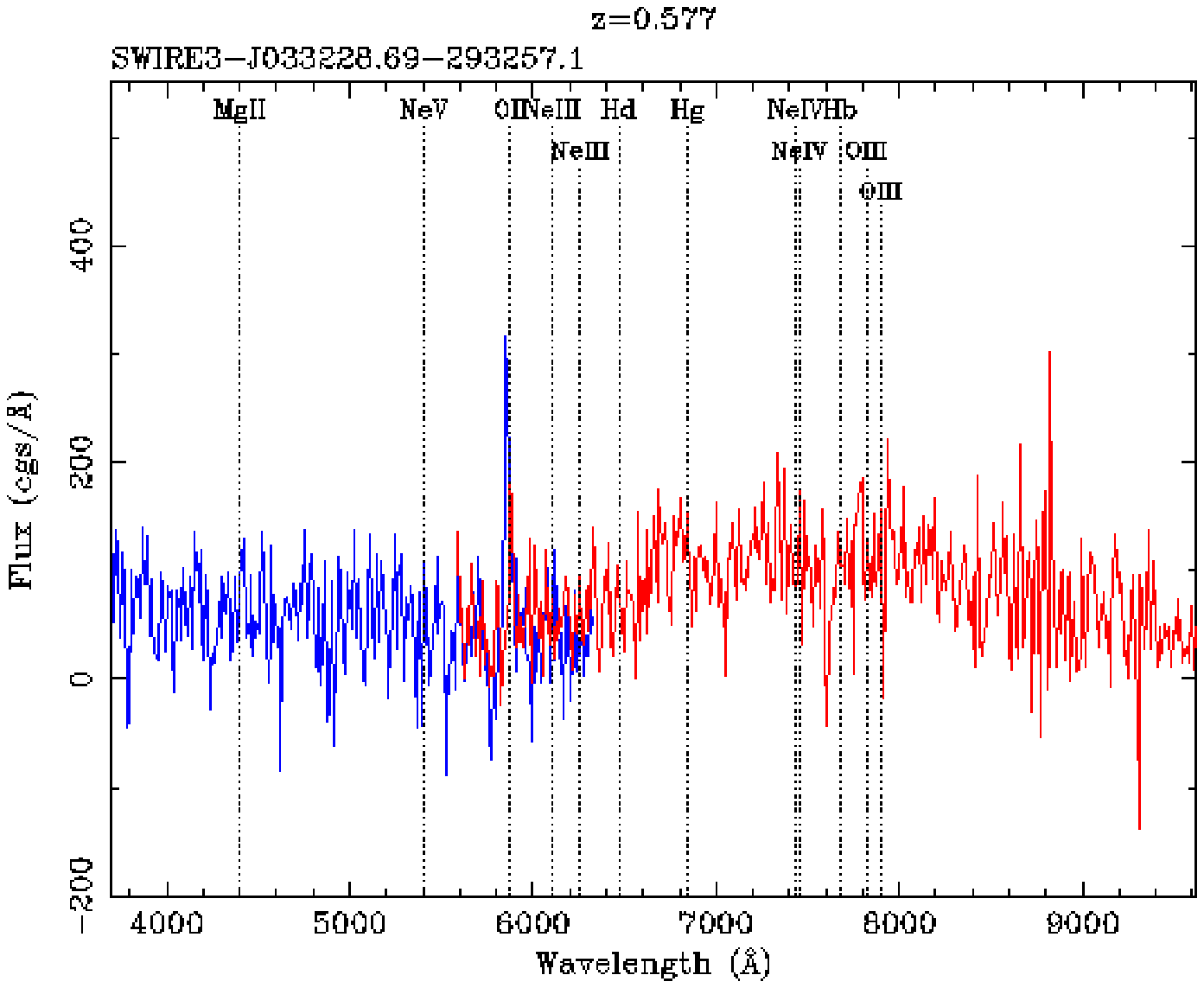}
\includegraphics[scale=0.55]{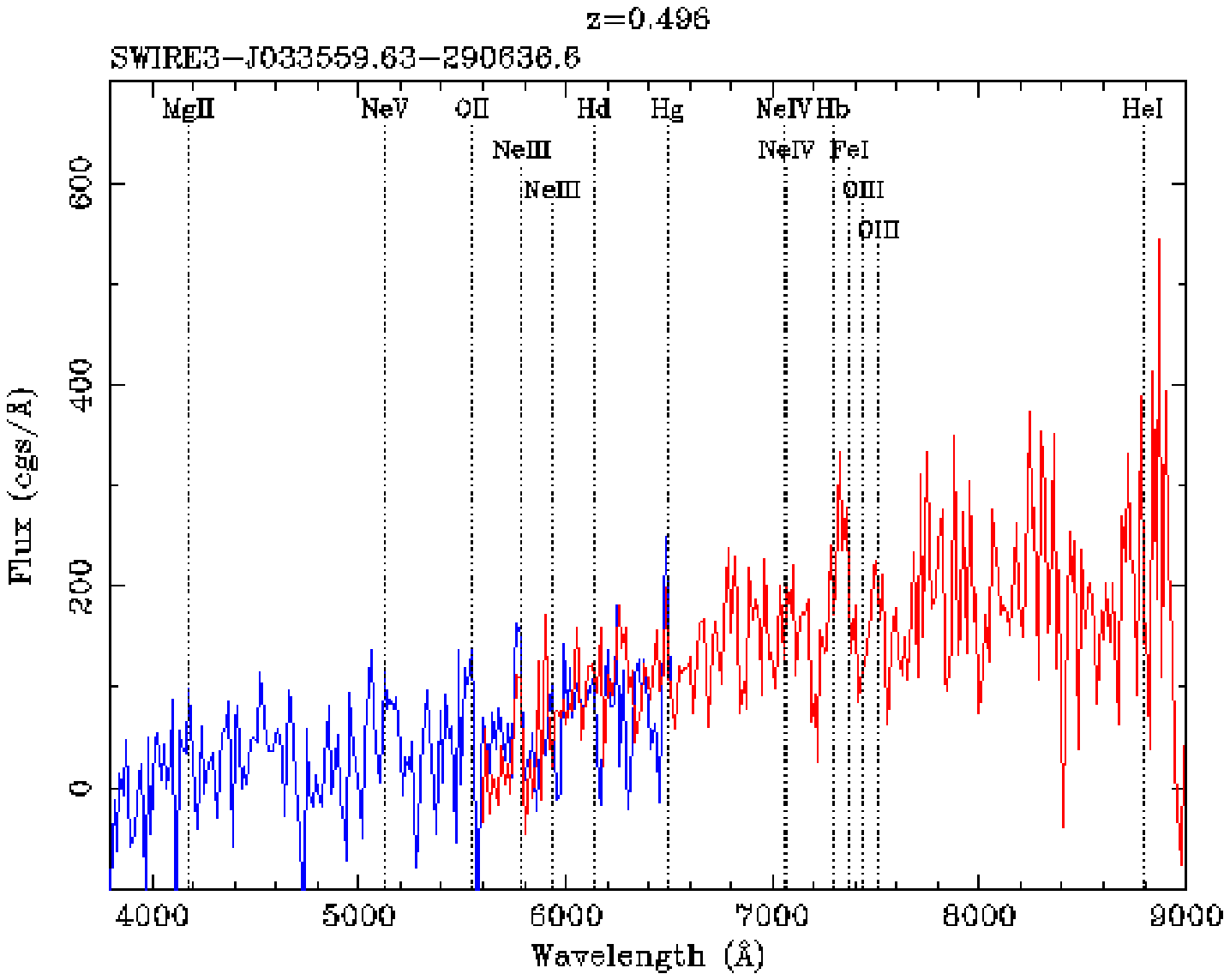}

\vspace{5pt}

\includegraphics[scale=0.55]{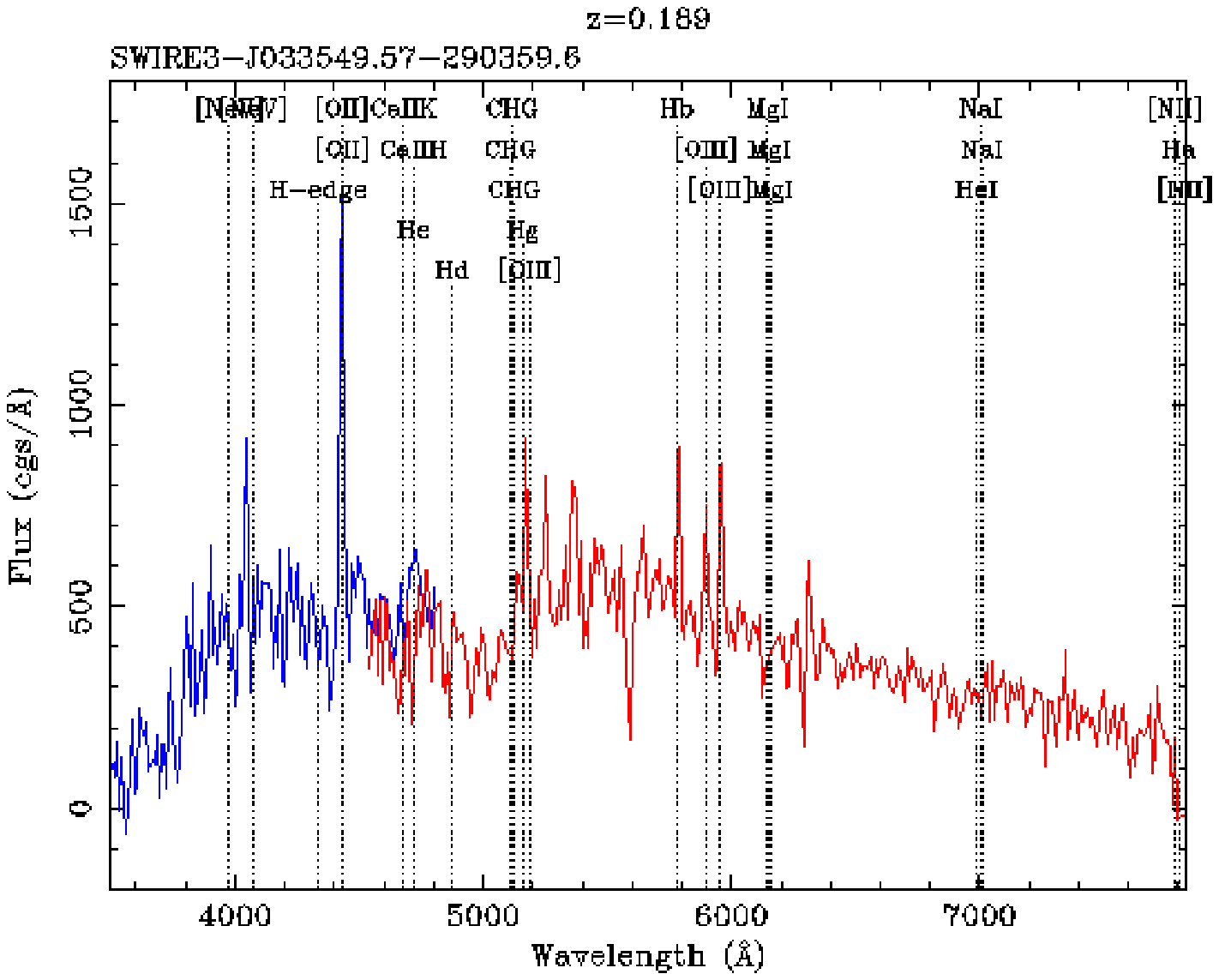}

\end{center}
\caption{Optical spectra of the rest 9 objects of our sample.}
\end{figure*}

\begin{table*}
\begin{minipage}{8.5in}

\caption{Parameters for the 34 EFOSC2 sources of our sample. (Luminosities and dust masses are in $log_{10}$ solar units)}
\tiny

\begin{tabular}{| c |c| c |c | c | c | c | c | c | c | c | c | c | c|}
\hline
\hline

Object		 		   &	RA   & 	DEC   &  {\it z}\footnote{Spectroscopic redshift} &  $L_{cirr}$\footnote{Bolometric luminosity in cirrus component} &  $L_{M82}$\footnote{Bolometric luminosity in starburst (M82 type) component} &  $L_{A220}$\footnote{Bolometric luminosity in starburst (A220 type) component}  &$L_{tor}$\footnote{Bolometric luminosity in torrus component} & $L_{ir}$\footnote{Bolometric infrared luminosity (1-1000$\mu$m) ($H_{0}=72km s^{-1} Mpc^{-1}$, $\lambda=0.7$)} & $L_{opt}$\footnote{Bolometric optical luminosity} & temp\footnote{Optical template type} & $A_v$\footnote{Visual extinction in magnitudes}  &	$M_{dust}$\footnote{Dust mass} & flag\footnote{Quality flag of spectroscopic redshift (quality A = multiple detections of strong features, quality B = redshift based on single emission line, \\ or based entirely on several weak, possibly spurious features)}\\
\hline
\hline
SWIRE3-J033531.60-283027.9 & 53.8816 & -28.5077 &  0.076  & 			   &			    & 		9.34		&		   &	 9.34 & 	9.59  & 	Scd  & 1.3 &	5.98 & B  \\
SWIRE3-J033532.56-283051.1 & 53.8856 & -28.5142 &  0.077  &  		 	   & 		8.61	& 					&		   &	 8.61 &		9.23  &		Sdm  & 0.0 &	4.65 & A  \\
SWIRE3-J033152.82-290647.1 & 52.9701 & -29.1131 &  0.079  & 			   & 		8.72	& 					&	 7.70  &	 8.76 &	    9.26  &		Scd  & 1.2 &	4.77 & A \\
SWIRE3-J033536.42-283309.5 & 53.9017 & -28.5526 &  0.105  & 		10.05  & 				&					&		   &	10.05 &	   10.31  & 	E 	 & 0.0 &	7.16 & A  \\
SWIRE3-J033148.59-290643.1 & 52.9524 & -29.1119 &  0.119  & 			   & 		10.88   & 					& 		   & 	10.88 &	   11.16  &		Scd  & 0.8 &	6.92 & A \\
SWIRE3-J033535.68-284714.2 & 53.8986 & -28.7873 &  0.123  & 		 9.49  & 		9.35	& 					& 		   & 	 9.72 &		9.99  &     sb   & 0.0 &	6.63 & A \\
SWIRE3-J033231.31-293452.9 & 53.1304 & -29.5814 &  0.158  & 		 9.62  & 		9.02	& 					& 		   &	 9.72 &    10.48  &		Scd  & 0.0 &	6.74 & B \\
SWIRE3-J033142.11-290551.9 & 52.9254 & -29.0977 &  0.158  &  		10.05  & 		8.93	& 					& 		   &    10.08 &    10.43  &     Scd  & 0.2 &	7.16 & A \\
SWIRE3-J033549.57-290359.6 & 53.9565 & -29.0665 &  0.189  & 		10.24  & 		10.55	& 					& 		   &    10.72 &    10.55  &     sb   & 0.7 &	7.42 & A \\
SWIRE3-J033552.63-290535.6 & 53.9693 & -29.0932 &  0.189  & 		10.52  & 		9.38 	& 				    & 		   & 	10.55 &    10.60  &		Scd  & 0.7 &	7.63 & B \\
SWIRE3-J033555.43-290758.9 & 53.9810 & -29.1330 &  0.190  & 		10.43  & 				& 				    &  		   &    10.43 &    10.51  &     Scd  & 0.5 &	7.54 & A \\
SWIRE3-J033551.15-290735.6 & 53.9631 & -29.1266 &  0.195  &  		 9.72  & 		9.38    & 					& 		   &     9.89 &    10.42  &     Scd  & 0.0 &	6.85 & A \\
SWIRE3-J033523.29-284827.2 & 53.8470 & -28.8075 &  0.195  & 			   & 	   10.93    & 	   				&	10.68  &	11.12 &    10.64  &     Sbc  & 0.0 &	6.97 & A \\
SWIRE3-J033008.28-285400.3 & 52.5344 & -28.9000 &  0.217  & 		10.53  & 			    & 				    & 		   &    10.53 &    11.01  &     sb   & 0.25&	7.64 & A \\
SWIRE3-J033558.74-290742.4 & 53.9947 & -29.1284 &  0.222  & 		10.48  & 	   10.52    & 					& 		   &    10.80 &     9.95  &     Sab  & 0.0 &	7.63 & A \\
SWIRE3-J033529.98-283344.4 & 53.8749 & -28.5623 &  0.225  & 		10.50  & 	   10.15	&					&		   &    10.66 &    11.11  &	    sb   & 0.15&	7.63 & A \\
SWIRE3-J033145.66-290728.0 & 52.9403 & -29.1244 &  0.245  & 			   & 	   			& 		10.76		&		   &    10.76 &    10.34  &     Scd  & 0.4 &	7.90 & A \\
SWIRE3-J033529.55-284559.2 & 53.8731 & -28.7664 &  0.297  & 			   & 	   11.20    & 					& 		   &    11.20 &    11.56  &     Sbc  & 0.0 &	7.24 & B \\
SWIRE3-J033534.18-283118.9 & 53.8924 & -28.5219 &  0.320  &				   & 	   10.16    & 				    &    9.96  &    10.37 &     9.77  &     QSO  & 0.3 &	6.20 & A \\
SWIRE3-J033016.50-285040.8 & 52.5688 & -28.8447 &  0.358  & 			   & 	   10.38    & 				    & 		   &    10.38 &    10.34  &     Scd  & 0.3 &	6.42 & A \\
SWIRE3-J033151.89-290606.1 & 52.9662 & -29.1016 &  0.372  & 			   & 	   10.68 	& 				    &  		   &    10.68 &    10.83  &     Sbc  & 0.0 &	6.72 & B \\
SWIRE3-J033227.95-293122.6 & 53.1164 & -29.5229 &  0.374  & 		10.75  & 	   10.87    &				    & 		   &    11.11 &    10.83  &     Scd  & 0.0 &	7.90 & B \\
SWIRE3-J033528.00-284500.3 & 53.8667 & -28.7501 &  0.404  & 			   &			    & 	   11.23 	    & 		   &	11.23 &    11.18  &     Scd  & 0.0 &	7.87 & A \\
SWIRE3-J033233.17-293004.8 & 53.1381 & -29.5013 &  0.453  & 			   & 			    &      11.56		&		   &    11.56 &    10.78  &     QSO  & 0.6 &	8.20 & A \\
SWIRE3-J033533.66-284620.5 & 53.8903 & -28.7724 &  0.453  & 			   & 	   10.03    &				    &   	   &    10.03 &    10.81  &     Scd  & 0.15&	6.07 & B \\
SWIRE3-J033559.63-290636.6 & 53.9985 & -29.1102 &  0.496  & 			   &       10.50	& 				    &   10.18  &    10.67 &    10.72  &     E    & 0.0 &	6.54 & B \\
SWIRE3-J033234.23-293450.8 & 53.1426 & -29.5808 &  0.529  & 			   & 	   12.11 	&				    & 		   &    12.11 &    11.45  &     Scd  & 0.0 &	8.15 & A \\
SWIRE3-J033228.69-293257.1 & 53.1195 & -29.5492 &  0.577  & 			   &	   10.82    & 					& 		   &    10.82 &    11.14  &     Scd  & 0.25&	6.86 & B \\
SWIRE3-J033541.23-283414.0 & 53.9217 & -28.5705 &  0.579  & 			   &       12.40    & 				    &   11.41  &    12.44 &    11.09  &     Scd  & 0.0 &	8.44 & A \\
SWIRE3-J033537.46-284354.0 & 53.9061 & -28.7317 &  0.769  & 			   &       11.64    &                   &          &    11.64 &    11.53  &     Scd  & 0.3 &	7.68 & B \\
SWIRE3-J033532.19-284801.1 & 53.8840 & -28.8003 &  0.807  & 			   &	   12.89    & 				    &   12.23  &  	12.98 &	   11.44  &	    E	 & 0.0 &	8.93 & B \\
SWIRE3-J033220.88-293140.5 & 53.0870 & -29.5279 &  1.180  & 			   & 				& 				    &   12.20  &  	12.20 &    11.92  &     QSO  & 0.0 &		 & A \\
SWIRE3-J033528.91-283203.6 & 53.8704 & -28.5343 &  1.962  & 			   & 			    & 				    &   12.53  &  	12.53 &    12.69  &     QSO  & 0.0 &		 & A \\
SWIRE3-J033144.54-290505.6 & 52.9356 & -29.0848 &  3.200  & 			   &       14.16    & 				    &   12.21  & 	14.16 &    12.01  &     QSO  & 0.0 &	10.20& A \\
\hline
\hline

\end{tabular}
\end{minipage}
\end{table*}

\bsp

\label{lastpage}

\end{document}